\documentclass[final]{siamltex}
\usepackage{graphicx}
\usepackage{psfrag}
\usepackage{amsfonts}
\usepackage{amssymb}
\usepackage{amscd}
\usepackage[mathscr]{euscript}
\usepackage{amsmath,latexsym,amsbsy}
\usepackage{enumerate}
\usepackage{epstopdf}
\usepackage{color}
\usepackage{bm}
\usepackage[square,comma,numbers,sort&compress]{natbib}

\newcommand*{\qed}{\hfill\ensuremath{\square}}%

\newtheorem{remark}{Remark}[section]

\def\R{\mathbb R}
\def\N{\mathbb N}
\def\ve{\varepsilon}

\def\oe{\overline}

\def\ve{\varepsilon}
\def\dc{\partial}
\def\bc{\begin{center}}
\def\ec{\end{center}}
\def\bt{\begin{tabular}}
\def\et{\end{tabular}}

\def\R{{\mathbb R}}

\def\y{{\bm{y}}}

\def\B{{\mathcal B}}

\def\ctan{{\rm ctan}}
\def\pa{{\partial\Omega}}
\def\a{{\mathfrak a}}

\def\D{{\mathcal D}}
\def\I{{\mathcal I}}

\def\diam{{\rm diam}}

\def\S{{\mathcal S}}     % for cross-section

\title{Mode matching methods for \\ spectral and scattering problems}

\author{A. Delitsyn\footnotemark[1] \ \footnotemark[2] 
\and D.~S. Grebenkov \footnotemark[3] \ \footnotemark[4] \ \footnotemark[5]}

\begin{document}
\maketitle

\renewcommand{\thefootnote}{\fnsymbol{footnote}}
\footnotetext[1]{Kharkevich Institute for Information Transmission Problems of RAN, Moscow, Russia}
\footnotetext[2]{The Main Research and Testing Robotics Center of the Ministry of Defense of Russia, Moscow, Russia}
\footnotetext[3]{Laboratoire de Physique de la Mati\`ere Condens\'ee, CNRS -- Ecole Polytechnique, University Paris-Saclay, 91128 Palaiseau, France}
\footnotetext[4]{Interdisciplinary Scientific Center Poncelet (ISCP), International Joint Research Unit (UMI 2615 CNRS/ IUM/ IITP RAS/ 
Steklov MI RAS/ Skoltech/ HSE), Bolshoy Vlasyevskiy Pereulok 11, 119002 Moscow, Russia}
\footnotetext[5]{Corresponding author: denis.grebenkov@polytechnique.edu}
\renewcommand{\thefootnote}{\arabic{footnote}}

\date{Received: \today / Revised version: }

\begin{abstract}
We present several applications of mode matching methods in spectral
and scattering problems.  First, we consider the eigenvalue problem
for the Dirichlet Laplacian in a finite cylindrical domain that is
split into two subdomains by a ``perforated'' barrier.  We prove that
the first eigenfunction is localized in the larger subdomain, i.e.,
its $L_2$ norm in the smaller subdomain can be made arbitrarily small
by setting the diameter of the ``holes'' in the barrier small enough.
This result extends the well known localization of Laplacian
eigenfunctions in dumbbell domains.  We also discuss an extension to
noncylindrical domains with radial symmetry.  Second, we study a
scattering problem in an infinite cylindrical domain with two
identical perforated barriers.  If the holes are small, there exists a
low frequency at which an incident wave is fully transmitted through
both barriers.  This result is counter-intuitive as a single barrier
with the same holes would fully reflect incident waves with low
frequences.
\end{abstract}



% 35J05 - Laplacian operator, reduced wave equation (Helmholtz equation), Poisson equation
% 35Pxx - Spectral theory and eigenvalue problems
% 51Pxx - Geometry and physics
% 33C10   	Bessel and Airy functions, cylinder functions, ${}_0F_1$

\today

\pagestyle{myheadings}
\thispagestyle{plain}

\section{Introduction}

Mode matching is a classical powerful method for the analysis of
spectral and scattering problems
\cite{Mittra,Levin,Marcuvitz,Weinstein,Collin}.  The main idea
of the method consists in decomposing a domain into ``basic''
subdomains, in which the underlying spectral or scattering problem can
be solved explicitly, and then matching the analytical solutions at
``junctions'' between subdomains.  This matching leads to functional
equations at the junctions and thus reduces the dimensionality of the
problem.  Such a dimensionality reduction is similar, to some extent,
to that in the potential theory when searching for a solution of the
Laplace equation in the bulk is reduced to finding an appropriate
``charge density'' on the boundary.  Although the resulting integral
or functional equations are in general more difficult to handle than
that of the original problem, they are efficient for obtaining
analytical estimates and numerical solutions.

\begin{figure}
\begin{center}
\includegraphics[width=120mm]{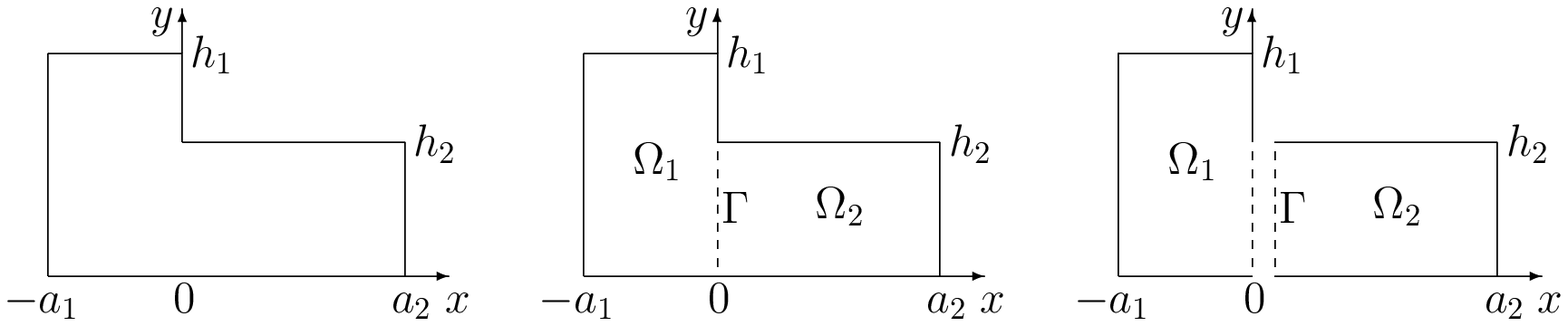} % Lshape2.eps}
\end{center}
\caption{
An L-shaped domain is decomposed into two rectangular subdomains
$\Omega_1$ and $\Omega_2$, in which explicit solutions are found and
then matched at the inner interface $\Gamma$. }
\label{fig:Lshape2}
\end{figure}

To illustrate the main idea, let us consider the eigenvalue problem
for the Dirichlet Laplacian in a planar L-shape domain
(Fig. \ref{fig:Lshape2}):
\begin{equation}
\label{eq:problem0}
-\Delta u = \lambda u \quad \mbox{ in } \Omega, \qquad u_{|\partial\Omega} = 0 .
\end{equation}
This domain can be naturally decomposed into two rectangular
subdomains $\Omega_1 = (-a_1,0)\times (0,h_1)$ and $\Omega_2 =
(0,a_2)\times (0,h_2)$.  Without loss of generality, we assume $h_1
\geq h_2$.  For each of these subdomains, one can explicitly write a
general solution of the equation in (\ref{eq:problem0}) due to a
separation of variables in perpendicular directions $x$ and $y$.  For
instance, one has in the rectangle $\Omega_1$
\begin{equation}
\label{eq:u_intro1}
u_1(x,y) = \sum\limits_{n=1}^\infty c_{n,1} \sin(\pi n y/h_1) \, \sinh(\gamma_{n,1} (a_1+x))  \qquad (x,y) \in \Omega_1, 
\end{equation}
where $\gamma_{n,1} = \sqrt{\pi^2 n^2/h_1^2 - \lambda}$ ensures that
each term satisfies (\ref{eq:problem0}).  The sine functions are
chosen to fulfill the Dirichlet boundary condition on the horizontal
edges of $\Omega_1$, while $\sinh(\gamma_{n,1} (a_1+x))$ vanishes on
the vertical edge at $x = -a_1$.  The coefficients $c_{n,1}$ are fixed
by the restriction $u_{1|\Gamma}$ of $u_1$ on the matching region
$\Gamma = (0,h_2)$ at $x = 0$.  In fact, multiplying
(\ref{eq:u_intro1}) by $\sin (\pi k y/h_1)$ and integrating over $y$
from $0$ to $h_1$, the coefficients $c_{n,1}$ are expressed through
$u_{1|\Gamma}$, and thus
\begin{equation}
\label{eq:u_intro2}
u_1(x,y) = \frac{2}{h_1} \sum\limits_{n=1}^\infty \bigl(u_{1|\Gamma}, \sin(\pi ny/h_1)\bigr)_{L_2(\Gamma)} \sin(\pi n y/h_1) 
\frac{\sinh(\gamma_{n,1} (a_1+x))}{\sinh(\gamma_{n,1} a_1)}  \, ,
\end{equation}
where $(\cdot,\cdot)_{L_2(\Gamma)}$ is the scalar product in the
$L_2(\Gamma)$ space.  Here we also used the Dirichlet boundary
condition on the vertical edge $\{0\}\times [h_2,h_1]$ to replace the
scalar product in $L_2(0,h_1)$ by that in $L_2(0,h_2) = L_2(\Gamma)$.
One can see that the solution of the equation (\ref{eq:problem0}) in
the subdomain $\Omega_1$ is fully determined by $\lambda$ and
$u_{1|\Gamma}$, which are yet unknown at this stage.

Similarly, one can write explicitly the solution $u_2$ in $\Omega_2$:
\begin{equation}
\label{eq:u_intro3}
u_2(x,y) = \frac{2}{h_2}\sum\limits_{n=1}^\infty \bigl(u_{2|\Gamma}, \sin(\pi ny/h_2)\bigr)_{L_2(\Gamma)} \sin(\pi n y/h_2) 
\frac{\sinh(\gamma_{n,2} (a_2-x))}{\sinh(\gamma_{n,2} a_2)}  \, ,
\end{equation}
where $\gamma_{n,2} = \sqrt{\pi^2 n^2/h_2^2 - \lambda}$.  Since the
solution $u$ is analytic in the whole domain $\Omega$, $u$ and its
derivative should be continuous at the matching region $\Gamma$:
\begin{equation}
u_{1|\Gamma} = u_{2|\Gamma} = u_{|\Gamma}  \, ,  \qquad 
\left.\frac{\partial u_1}{\partial x}\right|_{\Gamma} = \left.\frac{\partial u_2}{\partial x}\right|_{\Gamma} .
\end{equation}
Due to the explicit form (\ref{eq:u_intro2}, \ref{eq:u_intro3}) of the
solutions $u_1$ and $u_2$, the equality of the derivatives can be
written as
\begin{equation}
T_\lambda \, u_{|\Gamma} = 0   \qquad y \in \Gamma,
\end{equation}
where the auxiliary pseudo-differential operator $T_\lambda$ acts on a
function $v\in H^{\frac12}(\Gamma)$ (see rigorous definitions in
Sec. \ref{sec:1BD_intro}) as
\begin{equation}
\begin{split}
T_\lambda v & = \frac{2}{h_1} \sum\limits_{n=1}^\infty \gamma_{n,1} \coth(\gamma_{n,1} a_1)\, 
\bigl(v, \sin(\pi ny/h_1)\bigr)_{L_2(\Gamma)} \, \sin(\pi n y/h_1)  \\
& - \frac{2}{h_2}\sum\limits_{n=1}^\infty \gamma_{n,2} \coth(\gamma_{n,2} a_2)\, 
\bigl(v, \sin(\pi ny/h_2)\bigr)_{L_2(\Gamma)} \, \sin(\pi n y/h_2)   \,.\\
\end{split}
\end{equation}
The original eigenvalue problem for the Laplace operator in the
L-shaped domain $\Omega$ is thus reduced to a generalized eigenvalue
problem for the operator $T_\lambda$, with the advantage of the
reduced dimensionality, from a planar domain to an interval.  We have
earlier applied this technique to investigate trapped modes in finite
quantum waveguides \cite{Delitsyn12a}.

In this paper, we illustrate how mode matching methods can be used for
investigating spectral and scattering properties in various domains,
in particular, for deriving estimates for eigenvalues and
eigenfunctions.  More precisely, we address three problems:

(i) In Sec. \ref{sec:1BD}, we consider a finite cylinder $\Omega_0 =
[-a_1,a_2]\times \S \subset \R^{d+1}$ of a bounded connected
cross-section $\S \subset\R^d$ with a piecewise smooth boundary
$\partial\S$ (Fig. \ref{fig:domain}).  The cylinder is split into two
subdomains by a ``perforated'' barrier $\B \subset \S$ at $x = 0$,
i.e.  we consider the domain $\Omega = \Omega_0 \backslash (\{0\}
\times \B)$.  If $\B = \S$, the barrier separates $\Omega_0$ into two
disconnected subdomains, in which case the spectral analysis can be
done separately for each subdomain.  When $\B \ne \S$, an opening
region $\Gamma = \S\backslash \B$ (``holes'' in the barrier) connects
two subdomains.  When the diameter of the opening region,
$\diam\{\Gamma\}$, is small enough, we show that the first Dirichlet
eigenfunction $u$ is ``localized'' in a larger subdomain, i.e., the
$L_2$-norm of $u$ in the smaller subdomain vanishes as
$\diam\{\Gamma\}\to 0$.  This localization phenomenon resembles the
asymptotic behavior of Dirichlet eigenfunctions in dumbbell domains,
i.e., when two subdomains are connected by a narrow ``channel'' (see a
review \cite{Grebenkov13}).  When the width of the channel vanishes,
the eigenfunctions become localized in either of subdomains.  We
emphasize however that most of formerly used asymptotic techniques
(e.g., see
\cite{Raugel,Jimbo09,Daners03,Felli13,Arrieta95,Arrieta95b,Jimbo89,Hempel91,Jimbo92,Jimbo93,Brown95,Gadyl'shin93,Gadyl'shin94,Gadyl'shin05,Beale73,Melrose})
would fail in our case with no channel.  In fact, these former studies
dealt with the width-over-length ratio of the channel as a small
parameter that is not applicable in our situation as the channel
length is zero.  To our knowledge, we present the first rigorous proof
of localization in the case with no channel.  We discuss sufficient
conditions for localization.  A similar behavior can be observed for
other Dirichlet eigenfunctions, under stronger assumptions.  The
practical relevance of geometrically localized eigenmodes and their
physical applications were discussed in
\cite{Sapoval91,Even99,Felix07,Heilman10}.

(ii) In Sec. \ref{sec:2BD}, we consider a scattering problem in an
infinite cylinder $\Omega_0 = \R \times \S \subset \R^{d+1}$ of a
bounded cross-section $\S\subset\R^d$ with a piecewise smooth boundary
$\partial\S$.  The wave propagation in such waveguides and related
problems have been thoroughly investigated (see
\cite{Grebenkov13,Parker67,Ursell87,Ursell91,Evans92,Evans94,Exner96,Bulla97,Linton07,Hein08}
and references therein).  If the cylinder is blocked with a single
barrier $\B\subset \S$ with small holes, an incident wave is fully
reflected if its frequency is not high enough for a wave to
``squeeze'' through small holes.  Intuitively, one might think that
putting two identical barriers (Fig. \ref{fig:domain_TB}) would
enhance this blocking effect.  We prove that, if the holes in the
barriers are small enough, there exists a frequency close to the
smallest eigenvalue in the half-domain between two barriers, at which
the incident wave is fully transmitted through both barriers.  This
counter-intuitive result may have some acoustic applications.

(iii) In Sec. \ref{sec:cipollino}, we return to the eigenvalue problem
for the Dirichlet Laplacian and show an application of the mode
matching method to noncylindrical domains.  As an example, we consider
the union of a disk of radius $R_1$ and a part of a circular sector of
angle $\phi_1$ between two circles of radii $R_1$ and $R_2$
(Fig. \ref{fig:domain3}).  Under the condition that the sector is thin
and long, we prove the existence of an eigenfunction which is
localized in the sector and negligible inside the disk.  In
particular, we establish the inequalities between geometric parameters
$R_1$, $R_2$ and $\phi_1$ to ensure the localization.

%%%%%%%%%%%%  ONE BARRIER   %%%%%%%%%%%%%%%

\section{Barrier-induced localization in a finite cylinder}
\label{sec:1BD}

\subsection{Formulation and preliminaries}
\label{sec:1BD_intro}

\begin{figure}
\begin{center}
\includegraphics[width=120mm]{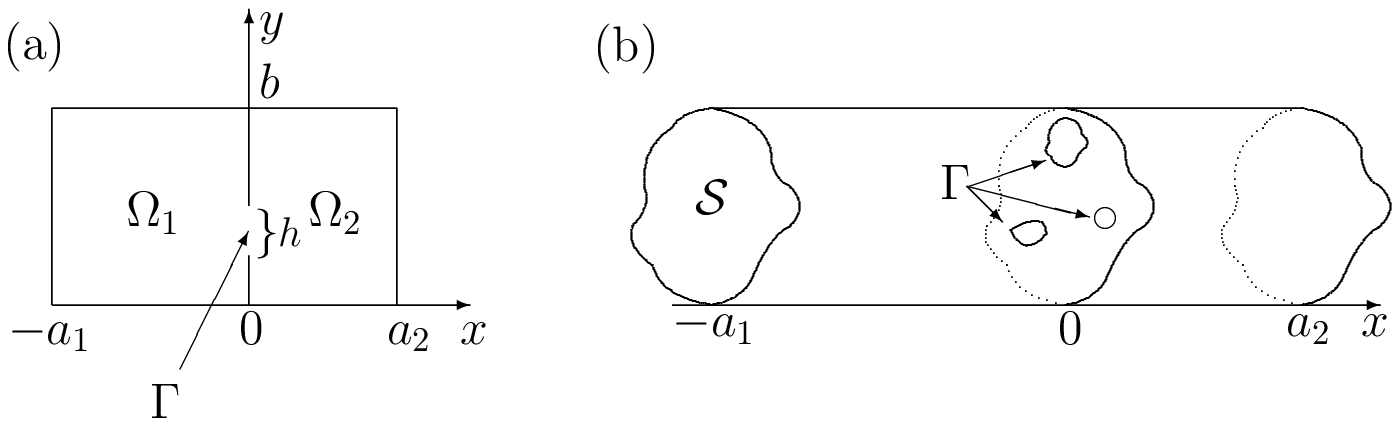}  % domain.eps}
\end{center}
\caption{
Illustration of cylindrical domains separated by a ``perforated''
barrier.  {\bf (a)} Two-dimensional case of a rectangle $\Omega_0 =
[-a_1,a_2]\times [0,b]$ separated by a vertical segment with a
``hole'' $\Gamma$ of width $h$.  {\bf (b)} Three-dimensional case of a
cylinder $\Omega_0 = [-a_1,a_2] \times \S$ of arbitrary cross-section
$\S$ separated by a barrier at $x =0$ with ``holes'' $\Gamma$.}
\label{fig:domain}
\end{figure}

Let $\Omega_0 = [-a_1,a_2]\times \S \subset \R^{d+1}$ be a finite
cylinder of a bounded connected cross-section $\S \subset \R^{d}$ with
a piecewise smooth boundary $\partial\S$ (Fig. \ref{fig:domain}).  For
a subdomain $\Gamma\subset \S$, let $\Omega = \Omega_0 \backslash
(\{0\}\times (\S \backslash \Gamma))$ be the cylinder without a cut at
$x = 0$ of the cross-sectional shape $\S \backslash \Gamma$.  The cut
$\{0\}\times (\S \backslash \Gamma)$ divides the domain $\Omega$ into
two cylindrical subdomains $\Omega_1$ and $\Omega_2$, connected
through $\Gamma$.  For instance, in two dimensions, one can take $\S =
[0,b]$ and $\Gamma = [h_1,h_2]$ (with $0 \leq h_1 < h_2 \leq b$) so
that $\Omega$ is a rectangle without a vertical slit of length $h =
h_2 - h_1$, as shown in Fig. \ref{fig:domain}(a).  We also denote by
$\S_x = \{x\} \times \S$ the cross-section at $x$.

We consider the Dirichlet eigenvalue problem in $\Omega$
\begin{equation}
\label{eq:eigen_1BD}
- \Delta u = \lambda u  \quad \textrm{in}~\Omega, \qquad  u_{|\partial \Omega} = 0 .
\end{equation}
We denote by $\nu_n$ and $\psi_n(\y)$ the Dirichlet eigenvalues and
$L_2(\S)$-normalized eigenfunctions of the Laplace operator
$\Delta_\perp$ in the cross-section $\S$:
\begin{equation}  \label{eq:eigen_perp_1BD}
- \Delta_\perp \psi_n(\y) = \nu_n \psi_n(\y) \quad (\y\in \S), \qquad  \psi_n(\y) = 0 \quad (\y\in \partial \S),
\end{equation}
where the eigenvalues are ordered: 
\begin{equation}  \label{eq:nu_def}
0 < \nu_1 < \nu_2 \leq \nu_2 \leq \ldots \nearrow +\infty .
\end{equation}

A general solution of (\ref{eq:eigen_1BD}) in each subdomain reads
\begin{eqnarray*}
u_1(x,\y) &\equiv&  u_{|\Omega_1} = \sum\limits_{n=1}^\infty c_{n,1}~ \psi_n(\y)~ \sinh(\gamma_n(a_1+x))  \qquad (-a_1 < x < 0), \\
u_2(x,\y) &\equiv&  u_{|\Omega_2} = \sum\limits_{n=1}^\infty c_{n,2}~ \psi_n(\y)~ \sinh(\gamma_n(a_2-x))  \qquad (0 < x < a_2),
\end{eqnarray*}
where 
\begin{equation}
\gamma_n = \sqrt{\nu_n - \lambda}
\end{equation}
can be either positive, or purely imaginary (in general, there can be
a finite number of purely imaginary $\gamma_n$ and infinitely many
real $\gamma_n$).  The coefficients $c_{n,1}$ and $c_{n,2}$ are
determined by multiplying $u_{1,2}(0,\y)$ at the matching
cross-section $x=0$ by $\psi_n(\y)$ and integrating over $\S_0$, from
which
\begin{equation}
\label{eq:1BD_u12}
\begin{split}
u_1(x,\y) &= \sum\limits_{n=1}^\infty b_n~ \psi_n(\y) ~ \frac{\sinh(\gamma_n(a_1+x))}{\sinh(\gamma_n a_1)}  \qquad (-a_1 < x < 0), \\
u_2(x,\y) &= \sum\limits_{n=1}^\infty b_n~ \psi_n(\y) ~ \frac{\sinh(\gamma_n(a_2-x))}{\sinh(\gamma_n a_2)}  \qquad (0 < x < a_2), \\
\end{split}
\end{equation}
where
\begin{equation}
\label{eq:bn_1BD}
b_n \equiv \bigl(u_{|\S_0}, \psi_n\bigr)_{L_2({\S_0})} = \bigl(u_{|\Gamma}, \psi_n\bigr)_{L_2(\Gamma)} ,
\end{equation}
with the conventional scalar product in $L_2(\Gamma)$:
\begin{equation}
\bigl(u, v\bigr)_{L_2(\Gamma)} = \int\limits_\Gamma d\y ~ u(\y) ~ v(\y)
\end{equation}
(since all considered operators are self-adjoint, we do not use
complex conjugate).  In the second identity in (\ref{eq:bn_1BD}), we
used the Dirichlet boundary condition on the barrier $\S_0\backslash
\Gamma$.  We see that the eigenfunction $u$ in the whole domain
$\Omega$ is fully determined by its restriction $u_{|\Gamma}$ to the
opening $\Gamma$.

Since the eigenfunction is analytic inside $\Omega$, the derivatives
of $u_1$ and $u_2$ with respect to $x$ should match on the opening
$\Gamma$:
\begin{equation}
\label{eq:matching_1BD}
\sum\limits_{n=1}^\infty b_n \gamma_n \bigl[\coth(\gamma_n a_1) + \coth(\gamma_n a_2)\bigr] \psi_n(\y) = 0  \quad (\y\in\Gamma)  .
\end{equation}
Multiplying this relation by a function $v \in H^{\frac12}(\Gamma)$
and integrating over $\Gamma$, one can introduce the associated
sesquilinear form $\a_\lambda(u,v)$:
\begin{equation}
\label{eq:1BD_adef}
\a_\lambda\bigl(u,v\bigr) = \sum\limits_{n=1}^\infty \gamma_n \bigl[\coth(\gamma_n a_1) + \coth(\gamma_n a_2)\bigr] \bigl(u, \psi_n \bigr)_{L_2(\Gamma)} 
\bigl(v, \psi_n \bigr)_{L_2(\Gamma)} ,
\end{equation}
where the Hilbert space $H^{\frac12}(\Gamma)$ is defined with the help
of eigenvalues $\nu_n$ and eigenfunctions $\psi_n(\y)$ of the
Dirichlet Laplacian $\Delta_\perp$ in the cross-section $\S$ as
\begin{equation}
H^{\frac12}(\Gamma) = \left\{ v\in L_2(\Gamma) ~:~ \sum\limits_{n=1}^\infty \sqrt{\nu_n} \, \bigl(v , \psi_n\bigr)^2_{L_2(\Gamma)} < +\infty \right\} .
\end{equation}
Note that this space equipped with the conventional scalar product:
\begin{equation}
\bigl(u,v\bigr)_{H^{\frac12}(\Gamma)} = \bigl(u,v\bigr)_{L_2(\Gamma)} + \sum\limits_{n=1}^\infty \sqrt{\nu_n} \, \bigl(u , \psi_n\bigr)_{L_2(\Gamma)}
\bigl(v , \psi_n\bigr)_{L_2(\Gamma)} .
\end{equation}

Using the sesquilinear form $\a_\lambda(u,v)$, one can understand the
matching condition (\ref{eq:matching_1BD}) in the weak sense as an
equation on $\lambda$ and $u_{|\Gamma} \in H^{\frac12}(\Gamma)$:
\begin{equation}
\label{eq:1BD_a0}
\a_\lambda\bigl(u_{|\Gamma},v\bigr) = 0    \qquad \forall~ v\in H^{\frac12}(\Gamma)
\end{equation}
(since this is a standard technique, we refer to textbooks
\cite{Kondrat'ev67,Ladyzhenskaya,Gilbarg,Lions,Grisvard1,Grisvard2} for
details).  Once a pair $\{\lambda, u_{|\Gamma}\} \in \R \times
H^{\frac12}(\Gamma)$ satisfying this equation for any $v\in
H^{\frac12}(\Gamma)$ is found, it fully determines the eigenfunction
$u \in H^1(\Omega)$ of the original eigenvalue problem
(\ref{eq:eigen_1BD}) in the whole domain $\Omega$ through
(\ref{eq:1BD_u12}).  In turn, this implies that $u
\in C^{\infty}(\Omega)$.  In other words, the mode matching method
allows one to reduce the original eigenvalue problem in the whole
domain $\Omega \subset \R^{d+1}$ to an equivalent problem
(\ref{eq:1BD_a0}) on the opening $\Gamma \subset \R^d$, thus reducing
the dimensionality of the problem.  More importantly, the reduced
problem allows one to derive various estimates on eigenvalues and
eigenfunctions.  In fact, setting $v = u_{\Gamma}$ yields the
dispersion relation
\begin{equation}
\label{eq:sum}
\a_\lambda\bigl(u_{|\Gamma}, u_{|\Gamma}\bigr) = \sum\limits_{n=1}^\infty b_n^2 \gamma_n \bigl[\coth(\gamma_n a_1) + \coth(\gamma_n a_2)\bigr] = 0  
\end{equation}
(with $b_n$ given by (\ref{eq:bn_1BD})), from which estimates on the
eigenvalue $\lambda$ can be derived (see below).  In turn, writing the
squared $L_2$-norm of the eigenfunction in an arbitrary cross-section
$\S_x$,
\begin{equation}
\label{eq:I_def}
I(x) \equiv ||u||^2_{L_2(\S_x)} = \int\limits_{\S} d\y ~ |u(x,\y)|^2  ,
\end{equation}
one gets
\begin{equation}
\label{eq:I_1BD}
I(x) = \begin{cases}
\displaystyle \sum\limits_{n=1}^\infty b_n^2 \frac{\sinh^2(\gamma_n(a_1+x))}{\sinh^2(\gamma_n a_1)}  \qquad (-a_1 < x < 0), \cr
\displaystyle \sum\limits_{n=1}^\infty b_n^2 \frac{\sinh^2(\gamma_n(a_2-x))}{\sinh^2(\gamma_n a_2)}  \qquad (0 < x < a_2),
\end{cases}
\end{equation}
and thus one can control the behavior of the eigenfunction $u$.

When the opening $\Gamma$ shrinks, the domain $\Omega$ is split into
two disjoint subdomains $\Omega_1$ and $\Omega_2$.  It is natural to
expect that each eigenvalue of the Dirichlet Laplacian in $\Omega$
converges to an eigenvalue of the Dirichlet Laplacian either in
$\Omega_1$, or in $\Omega_2$.  This statement will be proved in
Sec. \ref{sec:1BD_eigenvalues} and \ref{sec:1BD_eigenvalues1}.  The
behavior of eigenfunctions is more subtle.  If an eigenvalue in
$\Omega$ converges to a limit which is an eigenvalue in both
$\Omega_1$ and $\Omega_2$, the limiting eigenfunction is expected to
``live'' in both subdomains.  In turn, when the limit is an eigenvalue
of only one subdomain, one can expect that the limiting eigenfunction
will be localized in that subdomain.  In other words, for any integer
$N$, one expects the existence of a nonempty opening $\Gamma$ small
enough that at least $N$ eigenfunctions of the Dirichlet Laplacian in
$\Omega$ are localized in one subdomain (i.e., the $L_2$-norm of these
eigenfunctions in the other subdomain is smaller than a chosen $\ve
>0$).  Using the mode matching method, we prove a weaker form of these
yet conjectural statements.  We estimate the $L_2$ norm of an
eigenfunction $u$ in arbitrary cross-sections of $\Omega_1$ and
$\Omega_2$ and show that the ratio of these norms can be made
arbitrarily large under certain conditions.  The explicit geometric
conditions are obtained for the first eigenfunction in
Sec. \ref{sec:1BD_first}, while discussion about other eigenfunctions
is given in Sec. \ref{sec:1BD_other}.  Some numerical illustrations
are provided in Appendix \ref{sec:numerics}.

\subsection{Behavior of eigenvalues for the case $d > 1$}  \label{sec:1BD_eigenvalues}

We aim at showing that an eigenvalue of the Dirichlet Laplacian in
$\Omega$ is close to an eigenvalue of the Dirichlet Laplacian either
in $\Omega_1$, or in $\Omega_2$, for a small enough opening $\Gamma$.
In this subsection, we consider the case $d > 1$, which turns out to
be simpler and allows for more general statements.  The planar case
(with $d = 1$) will be treated separately in
Sec. \ref{sec:1BD_eigenvalues1}.

The proof relies on the general classical
\begin{lemma}  \label{lem:Av_eps}
Let $A$ be a self-adjoint operator with a discrete spectrum, and there
exist constants $\ve > 0$ and $\mu\in \R$ and a function $v$ from the
domain $\D_A$ of the operator $A$ such that
\begin{equation}
\label{eq:Av_eps}
\| A v - \mu v \|_{L_2} < \ve \| v\|_{L_2} .
\end{equation}
Then there exists an eigenvalue $\lambda$ of $A$ such that
\begin{equation}
|\lambda - \mu| < \ve .
\end{equation}
\end{lemma} \\
An elementary proof is reported in Appenix \ref{sec:proof_lemma} for
completeness.

The lemma states that if one finds an approximate eigenpair $\mu$ and
$v$ of the operator $A$, then there exists its true eigenvalue
$\lambda$ close to $\mu$.  Since we aim at proving the localization of
an eigenfunction in one subdomain, we expect that an appropriately
extended eigenpair in this subdomain can serve as $\mu$ and $v$ for
the whole domain.

\begin{theorem}  \label{lem:main_1BD}
Let $a_1$ and $a_2$ be strictly positive nonequal real numbers, $\S$
be a bounded domain in $\R^d$ with $d > 1$ and a piecewise smooth
boundary $\partial\S$, $\Gamma$ be an nonempty subset of $\S$, and
\begin{equation}
\Omega = ([-a_1,a_2]\times \S)\backslash (\{0\} \times \Gamma) , \quad
\Omega_1 = [-a_1,0]\times \S,  \quad \Omega_2 = [0,a_2] \times \S .
\end{equation}
Let $\mu$ be any eigenvalue of the Dirichlet Laplacian in the
subdomain $\Omega_2$.  Then for any $\ve > 0$, there exists $\delta >
0$ such that for any opening $\Gamma$ with $\diam\{\Gamma\} < \delta$,
there exists an eigenvalue $\lambda$ of the Dirichlet Laplacian in
$\Omega$ such that $|\lambda - \mu| < \ve$.  The same statement holds
for the subdomain $\Omega_1$.
\end{theorem} \\
{\bf Proof.}  We will prove the statement for the subdomain
$\Omega_2$.  The Dirichlet eigenvalues and eigenfunctions of the
Dirichlet Laplacian in $\Omega_2$ are
\begin{equation}  \label{eq:1BD_v}
\begin{array}{l l}
\mu_{n,k} = \pi^2 k^2/a_2^2 + \nu_n  \\  
v_{n,k} = \sqrt{2/a_2} \, \sin(\pi k x/a_2) \, \psi_n(\y)  \\ \end{array} \qquad (n,k= 1,2,\ldots) ,
\end{equation}
where $\nu_n$ and $\psi_n(\y)$ are the Dirichlet eigenpairs in the
cross-section $\S$, see (\ref{eq:eigen_perp_1BD}).

Let $\mu = \mu_{n,k}$ and $v = v_{n,k}$ for some $n,k$.  Let $2\delta$
be the diameter of the opening $\Gamma$, and $\y_{\Gamma} \in \S$ be
the center of a ball $B_{(0,\y_\Gamma)}(\delta) \subset \R^{d+1}$ of
radius $\delta$ that encloses $\Gamma$.  We introduce a cut-off
function $\bar \eta$ defined on $\Omega$ as
\begin{equation}
\bar\eta(x,\y) = \I_{\Omega_2}(x,\y) \, \eta\bigl(|(x,\y) - (0,\y_\Gamma)|/\delta\bigr) ,
\end{equation}
where $\I_{\Omega_2}(x,\y)$ is the indicator function of $\Omega_2$
(which is equal to $1$ inside $\Omega_2$ and $0$ otherwise), and
$\eta(r)$ is an analytic function on $\R_+$ which is 0 for $r < 1$ and
$1$ for $r > 2$.  In other words, $\bar\eta(x,\y)$ is zero outside
$\Omega_2$ and in a $\delta$-vicinity of the opening $\Gamma$, it
changes to $1$ in a thin spherical shell of width $\delta$, and it is
equal to $1$ in the remaining part of $\Omega_2$.

According to Lemma \ref{lem:Av_eps}, it is sufficient to check that
\begin{equation}
\label{eq:1BD_auxil7}
\| \Delta (\bar\eta\, v) + \mu \, \bar\eta\, v \|_{L_2(\Omega)} < \ve  \|\bar\eta\, v \|_{L_2(\Omega)} .
\end{equation}
We have
\begin{eqnarray}
\nonumber
\| \Delta (\bar\eta\, v) + \mu\, \bar\eta \, v \|^2_{L_2(\Omega)} 
&=& \| \bar\eta \underbrace{(\Delta v + \mu\, v)}_{=0} + 2 (\nabla \bar\eta \cdot \nabla v) + v \Delta \bar\eta \|^2_{L_2(\Omega)} \\
\label{eq:1BD_auxil23}
& \leq & 2 \| 2 (\nabla \bar\eta \cdot \nabla v) \|^2_{L_2(\Omega)} + 2 \| v \, \Delta \bar\eta \|^2_{L_2(\Omega)} ,
\end{eqnarray}
where $(\nabla \bar\eta \cdot \nabla v)$ is the scalar product between
two vectors in $\R^{d+1}$.  Due to the cut-off function $\bar\eta$,
one only needs to integrate over a part of the spherical shell around
the point $(0,\y_\Gamma)$:
\begin{equation}
Q_\delta = \{ (x,\y)\in \Omega_2 ~:~ \delta < |(x,\y) - (0,\y_\Gamma)| < 2\delta\} .  
\end{equation}
It is therefore convenient to introduce the spherical coordinates in
$\R^{d+1}$ centered at $(0,\y_\Gamma)$, with the North pole directed
along the positive $x$ axis.

For the first term in (\ref{eq:1BD_auxil23}), we have
\begin{equation}
\| (\nabla v \cdot \nabla \bar\eta)\|^2_{L_2(\Omega)} = \int\limits_{Q_\delta} dx \, d\y \,  |\bigl(\nabla v \cdot \nabla \bar\eta\bigr)|^2
= \int\limits_{Q_\delta} dx \, d\y \,  \biggl(\frac{\partial v}{\partial r} \, \frac{d\bar\eta(r/\delta)}{dr}\biggr)^2 ,
\end{equation}
because the function $\bar\eta$ varies only along the radial
direction.  Since $v$ is an analytic function inside $\Omega_2$,
$\partial v/\partial r$ is bounded over $\Omega_2$, so that
\begin{equation}
\| (\nabla v \cdot \nabla \bar\eta)\|^2_{L_2(\Omega)} \leq
\max\limits_{(x,\y)\in \Omega_2} \biggl\{ \biggl(\frac{\partial v}{\partial r}\biggr)^2 \biggr\}
\int\limits_{\delta}^{2\delta} dr ~ r^d \biggl(\frac{d\eta(r/\delta)}{dr}\biggr)^2  \int\limits_{x>0} d\Theta_d,
\end{equation}
where $d\Theta_d$ includes all angular coordinates taken over the half
of the sphere ($x > 0$).  Changing the integration variable, one gets
\begin{equation}  \label{eq:1BD_auxil33a}
\| (\nabla v \cdot \nabla \bar\eta)\|^2_{L_2(\Omega)} \leq C_1 \delta^{d-1} ,
\end{equation}
with a constant
\begin{equation}
C_1 = \max\limits_{(x,\y)\in \Omega_2} \biggl\{ \biggl(\frac{\partial v}{\partial r}\biggr)^2 \biggr\}
\int\limits_{1}^{2} dr ~ r^d \biggl(\frac{d \eta(r)}{dr}\biggr)^2  \int\limits_{x>0} d\Theta_d .
\end{equation}

For the second term in (\ref{eq:1BD_auxil23}), we have
\begin{eqnarray}
\nonumber
\|v\, \Delta \bar\eta \|^2_{L_2(\Omega)} &=& \int\limits_{Q_\delta} dx\, d\y \, (2/a_2) \,\sin^2(\pi n x/a_2)\, \psi_k^2(\y) \, (\Delta \bar\eta)^2 \\
&\leq& (2/a_2) \max\limits_{\y\in \S} \{ \psi_k^2(\y) \} \int\limits_{Q_\delta} dx \, d\y \, (\pi n x/a_2)^2\,  (\Delta \bar\eta)^2 , 
\end{eqnarray}
where we used the explicit form of the eigenfunction $v$ from
(\ref{eq:1BD_v}), the inequality $\sin x < x$ for $x > 0$, and the
boundedness of $\psi_k(\y)$ over a bounded cross-section $\S$ to get
rid off $\psi_k^2$.  Denoting $C_2 = (2/a_2) (\pi n/a_2)^2
\max\limits_{\y\in \S} \{ \psi_k^2(\y) \}$, we get
\begin{eqnarray}
\nonumber
\|v\, \Delta \bar\eta \|^2_{L_2(\Omega)} & \leq & C_2 \int\limits_\delta^{2\delta} dr~ r^{d} \int\limits_{x>0} d\Theta_d ~ r^2 \cos^2\theta  
\left(\frac{1}{r^{d}} \frac{d}{dr} r^{d} \frac{d}{dr} \eta(r/\delta) \right)^2 \\
& \leq & C_2 \delta^{d-1} \int\limits_1^{2} dr~ r^{d+2}  ~ 
\left(\frac{1}{r^{d}} \frac{d}{dr} r^{d} \frac{d}{dr} \eta(r) \right)^2 \int\limits_{x>0} d\Theta_d  , 
\end{eqnarray}
where $\theta$ is the azimuthal angle from the $x$ axis.  The
remaining integral is just a constant which depends on the particular
choice of the cut-off function $\eta(r)$.  We get thus
\begin{equation}  \label{eq:1BD_auxil22}
\|v\, \Delta \bar\eta \|^2_{L_2(\Omega)} \leq C_2' \delta^{d-1} ,
\end{equation}
with a new constant $C_2'$.

Combining inequalities (\ref{eq:1BD_auxil23}, \ref{eq:1BD_auxil33a},
\ref{eq:1BD_auxil22}), we obtain
\begin{equation}
\| \Delta (\bar\eta\, v) + \mu\, \bar\eta \, v \|^2_{L_2(\Omega)} \leq C \, \delta^{d-1} ,
\end{equation}
with a new constant $C$.

On the other hand,
\begin{eqnarray}
\nonumber
\| \bar\eta \, v \|^2_{L_2(\Omega)} &=& \int\limits_{\Omega_2} dx \, d\y \, \bar\eta^2 \, v^2 
= 1 - \int\limits_{B_{2\delta}} dx \, d\y \, (1-\bar\eta^2) \, v^2 \\
\label{eq:1BD_auxil33}
& \geq& 1 - \max\limits_{\Omega_2} \{ v^2 \} \int\limits_0^{2\delta} dr \, r^d \int\limits_{x>0} d\Theta_d \, \eta^2(r/\delta) = 1 - C_0 \delta^{d+1} , 
\end{eqnarray}
where $C_0$ is a constant, $B_{2\delta} = \{(x,\y)\in\Omega_2~:~
|(x,\y)-(0,\y_\Gamma)| < 2\delta\}$ is a half-ball of radius $2\delta$
centered at $(0,\y_\Gamma)$, and we used the $L_2(\Omega_2)$
normalization of $v$ and its boundedness.  In other words, for
$\delta$ small enough, the left-hand side of (\ref{eq:1BD_auxil33})
can be bounded from below by a strictly positive constant.  Recalling
that $2\delta$ is the diameter of $\Gamma$, we conclude that
inequality (\ref{eq:1BD_auxil7}) holds with $\ve = C'
(\diam\{\Gamma\})^{d-1}$, with some $C' > 0$.  If the diameter is
small enough, Lemma \ref{lem:Av_eps} implies the existence of an
eigenvalue $\lambda$ of the Dirichlet Laplacian in $\Omega$ close to
$\mu$, which is an eigenvalue in $\Omega_2$ that completes the proof.
\qed

\begin{remark}
In this proof, the half-diameter $\delta$ of the opening $\Gamma$
plays the role of a small parameter, whereas the geometric structure
of $\Gamma$ does not matter.  If $\Gamma$ is the union of a finite
number of disjoint ``holes'', the above estimates can be improved by
constructing cut-off functions around each ``hole''.  In this way, the
diameter of $\Gamma$ can be replaced by diameters of each ``hole''.
Note also that the proof is not applicable to a narrow but elongated
opening (e.g., $\Gamma = (0,h)\times (0,1/2)$ inside the square
cross-section $\S = (0,1)\times (0,1)$ with small $h$): even if the
Lebesgue measure of the opening can be arbitrarily small, its diameter
can remain large.  We expect that the theorem might be extended to
such situations but finer estimates are needed.
\end{remark}

\subsection{Behavior of the first eigenvalue for the case $d = 1$}
\label{sec:1BD_eigenvalues1}

The above proof is not applicable in the planar case (with $d = 1$).
For this reason, we provide another proof which is based on the
variational analysis of the modified eigenvalue problem with the
sesquilinear form $\a_\lambda(u,v)$.  This proof also serves us as an
illustration of advantages of mode matching methods.  For the sake of
simplicity, we only focus on the behavior of the first (smallest)
eigenvalue.

Without loss of generality, we assume that
\begin{equation}
\label{eq:1BD_assump1}
a_1 > a_2, 
\end{equation}
i.e, the subdomain $\Omega_1$ is larger than $\Omega_2$.  

\begin{lemma}[Domain monotonicity]
The domain monotonicity for the Dirichlet Laplacian implies the
following inequalities for the first eigenvalue $\lambda$ in $\Omega$
\begin{equation}
\nu_1 + \frac{\pi^2}{(a_1+a_2)^2} < \lambda < \nu_1 + \frac{\pi^2}{a_1^2} \,,
\end{equation}
where $\nu_1 + \pi^2/(a_1+a_2)^2$ is the smallest eigenvalue in
$[-a_1,a_2]\times \S$, $\nu_1 + \pi^2/a_1^2$ is the smallest
eigenvalue in $[-a_1,0]\times \S$, and $\nu_1$ is the smallest
eigenvalue in $\S$.
\end{lemma} \\
This lemma implies that $\gamma_1 = \sqrt{\nu_1 - \lambda}$ is purely
imaginary.  To ensure that the other $\gamma_n$ with $n\geq 2$ are
positive, we assume that
\begin{equation}
\label{eq:1BD_assump2}
a_1 \geq \frac{\pi}{\sqrt{\nu_2-\nu_1}} \,. 
\end{equation}

\begin{theorem}
\label{lem:eigen_1BD}
Let $a_1 \geq 1/\sqrt{3}$ and $0 < a_2 < a_1$ be two real numbers, $\S
= [0,1]$, $\Gamma$ be an nonempty subset of $\S$, and
\begin{equation}
\Omega = ([-a_1,a_2]\times \S)\backslash (\{0\} \times \Gamma) , \quad
\Omega_1 = [-a_1,0]\times \S,  \quad \Omega_2 = [0,a_2] \times \S .
\end{equation}
Let $\mu = \pi^2 + \pi^2/a_1^2$ be the smallest eigenvalue of the
Dirichlet Laplacian in the (larger) rectangle $\Omega_1$.  Then for
any $\ve > 0$, there exists $\delta > 0$ such that for any opening
$\Gamma$ with $\diam\{\Gamma\} = h < \delta$, there exists an
eigenvalue $\lambda$ of the Dirichlet Laplacian in $\Omega$ such that
$|\lambda - \mu| < \ve$.
\end{theorem} \\
{\bf Proof.}  As discussed earlier, the eigenfunction $u$ in the whole
domain is fully determined by its restriction on the opening $\Gamma$
that obeys (\ref{eq:1BD_a0}).  This equation can also be seen as an
eigenvalue problem
\begin{equation}
\a_\lambda\bigl(u_{|\Gamma},v\bigr) = \eta(\lambda) \bigl(u_{|\Gamma},v\bigr)_{L_2(\Gamma)}   \qquad \forall~ v\in H^{\frac 12}(\Gamma),
\end{equation}
where the eigenvalue $\eta(\lambda)$ depends on $\lambda$ as a
parameter.  The smallest eigenvalue can then be written as
\begin{equation}
\label{eq:mu1_1BD}
\eta_1(\lambda) = \inf\limits_{\substack{ v\in H^{\frac12}(\Gamma) \\ v \ne 0}} \biggl\{\frac{F(v)}{||v||^2_{L_2(\Gamma)}}   \biggr\},
\end{equation}
with
\begin{equation}
\label{eq:Fv_1BD}
F(v) = \sum\limits_{n=1}^\infty \gamma_n \bigl[\coth(\gamma_n a_1) + \coth(\gamma_n a_2)\bigr] \bigl(v, \psi_n \bigr)^2_{L_2(\Gamma)}.
\end{equation}
If $\eta_1(\lambda_c) = 0$ at some $\lambda_c$, then $\lambda_c$ is an
eigenvalue of the original problem.  We will prove that the zero
$\lambda_c$ of $\eta_1(\lambda)$ converges to $\mu$ as the opening
$\Gamma$ shrinks.

First, we rewrite the functional $F(v)$ under the assumptions
(\ref{eq:1BD_assump1}, \ref{eq:1BD_assump2}) ensuring that $\gamma_1$
is purely imaginary while $\gamma_n$ with $n \geq 2$ are positive:
\begin{eqnarray} \nonumber
F(v) & = & |\gamma_1| \bigl[\ctan(|\gamma_1| a_1) + \ctan(|\gamma_1| a_2)\bigr] \bigl(v, \psi_1 \bigr)^2_{L_2(\Gamma)}  \\
\label{eq:Fv_1BDa}
& + & \sum\limits_{n=2}^\infty \gamma_n \bigl[\coth(\gamma_n a_1) + \coth(\gamma_n a_2)\bigr] \bigl(v, \psi_n \bigr)^2_{L_2(\Gamma)} .
\end{eqnarray}

On one hand, an upper bound reads
\begin{eqnarray}  \nonumber
&& \sum\limits_{n=2}^\infty \gamma_n \bigl[\coth(\gamma_n a_1) + \coth(\gamma_n a_2)\bigr] \bigl(v, \psi_n \bigr)^2_{L_2(\Gamma)} \\
&&\leq C_1 \sum\limits_{n=2}^\infty \sqrt{\nu_n} \bigl(v, \psi_n \bigr)^2_{L_2(\Gamma)} 
 \leq C_2 \|v\|^2_{H^{\frac12}(\Gamma)} ,
\end{eqnarray}
because $\gamma_n = \sqrt{\nu_n - \lambda} \leq C_1'
\sqrt{\nu_n}$ for all $n\geq 2$.  As a consequence,
\begin{equation}
\eta_1(\lambda) \leq \frac{F(v)}{\|v\|^2_{L_2(\Gamma)}} \leq |\gamma_1| \bigl[\ctan(|\gamma_1| a_1) + \ctan(|\gamma_1| a_2)\bigr]
\frac{\bigl(v, \psi_1 \bigr)^2_{L_2(\Gamma)}}{\|v\|^2_{L_2(\Gamma)}}  + C_2 \frac{\|v\|^2_{H^{\frac12}(\Gamma)}}{\|v\|^2_{L_2(\Gamma)}},
\end{equation}
where $v$ can be any smooth function from $H^{\frac12}(\Gamma)$ that
vanishes at $\partial\Gamma$ and is not orthogonal to $\psi_1$, i.e.,
$\bigl(v, \psi_1 \bigr)_{L_2(\Gamma)} \ne 0$.  Since $\ctan(|\gamma_1|
a_1) \to -\infty$ as $|\gamma_1| a_1 \to \pi$, one gets {\it negative}
values $\eta_1(\lambda)$ for $\lambda$ approaching $\mu$.

On the other hand, for any fixed $\lambda$, we will show that
$\eta_1(\lambda)$ becomes positive as $\diam\{\Gamma\} \to 0$.  For
this purpose, we write
\begin{eqnarray} \nonumber
\eta_1(\lambda) & = & \inf\limits_{\substack{v\in H^{\frac 12}(\Gamma) \\ v\ne 0}} 
\left\{ \beta \frac{\bigl(v, \psi_1 \bigr)^2_{L_2(\Gamma)}}{||v||^2_{L_2(\Gamma)}} 
+ \frac{F_1(v)}{||v||^2_{L_2(\Gamma)}} \right\}, \\
\label{eq:1BD_auxil2}
& \geq & \inf\limits_{\substack{ v\in H^{\frac 12}(\Gamma) \\ v\ne 0}} 
\left\{ \beta \frac{\bigl(v, \psi_1 \bigr)^2_{L_2(\Gamma)}}{||v||^2_{L_2(\Gamma)}} \right\}
+ \inf\limits_{\substack{v\in H^{\frac 12}(\Gamma) \\ v\ne 0}} \left\{ \frac{F_1(v)}{||v||^2_{L_2(\Gamma)}} \right\}, 
\end{eqnarray}
with
\begin{equation}
\beta = |\gamma_1| \bigl[\ctan(|\gamma_1| a_1) + \ctan(|\gamma_1| a_2) - \coth(|\gamma_1| a_1) - \coth(|\gamma_1| a_2)\bigr]
\end{equation}
and
\begin{equation}
F_1(v) = \sum\limits_{n=1}^\infty |\gamma_n| \bigl[\coth(|\gamma_n| a_1) + \coth(|\gamma_n| a_2)\bigr]  \bigl(v, \psi_n \bigr)^2_{L_2(\Gamma)}.
\end{equation}
If $\beta \geq 0$, the first infimum in (\ref{eq:1BD_auxil2}) is
bounded from below by $0$.  If $\beta < 0$, the first infimum is
bounded from below as
\begin{equation}
\inf\limits_{\substack{v\in H^{\frac 12}(\Gamma) \\ v\ne 0}} \left\{ \beta \frac{\bigl(v, \psi_1 \bigr)^2_{L_2(\Gamma)}}{||v||^2_{L_2(\Gamma)}} \right\}
= - |\beta| \sup\limits_{\substack{v\in H^{\frac 12}(\Gamma) \\ v\ne 0}} 
\left\{ \frac{\bigl(v, \psi_1 \bigr)^2_{L_2(\Gamma)}}{||v||^2_{L_2(\Gamma)}} \right\} \geq - |\beta| ,
\end{equation}
because 
\begin{equation}
0\leq \frac{\bigl(v, \psi_n \bigr)^2_{L_2(\Gamma)}}{||v||^2_{L_2(\Gamma)}} \leq ||\psi_n||^2_{L_2(\Gamma)} \leq ||\psi_n||^2_{L_2(\S)} = 1 
\qquad (\forall ~ n \geq 1).
\end{equation}
We conclude that

\begin{equation}
\eta_1(\lambda) \geq \min\{\beta,0\} + \inf\limits_{\substack{v\in H^{\frac 12}(\Gamma) \\ v\ne 0}} \left\{ \frac{F_1(v)}{||v||^2_{L_2(\Gamma)}} \right\} \,.
\end{equation}
Since $\beta$ does not depend on $\Gamma$, it remains to check that
the second term diverges as $\diam\{\Gamma\} \to 0$ that would ensure
positive values for $\eta_1(\lambda)$.  Since
\begin{equation}
F_1(v) \geq C F_0(v), \qquad F_0(v) = \sum\limits_{n=1}^\infty \sqrt{\nu_n} ~ \bigl(v, \psi_n \bigr)^2_{L_2(\Gamma)}
\end{equation}
for some constant $C > 0$, it is enough to prove that
\begin{equation}
\inf\limits_{\substack{v\in H^{\frac 12}(\Gamma) \\ v\ne 0}} \biggl\{ \frac{F_0(v)}{||v||_{L_2(\Gamma)}} \biggr\}  
\xrightarrow[\diam\{\Gamma\}\to 0]{} +\infty  .
\end{equation}
This is proved in Lemma \ref{lem:infimum}.  When the opening $\Gamma$
shrinks, $\eta_1(\lambda)$ becomes {\it positive}.

Since $\eta_1(\lambda)$ is a continuous function of $\lambda$ (see
Lemma \ref{lem:1BD_continuity}), there should exist a value
$\lambda_c$ at which $\eta_1(\lambda_c) = 0$.  This is the smallest
eigenvalue of the original problem which is close to $\mu$.  This
completes the proof of the theorem.  \qed

\begin{lemma}  \label{lem:infimum}
For an nonempty set $\Gamma \subset [0,1]$, we have
\begin{equation}
I(\Gamma) \equiv \inf\limits_{\substack{v\in H^{\frac 12}(\Gamma) \\ v\ne 0}} \left\{ \frac{\sum\limits_{n=1}^\infty n 
\bigl(v, \sin(\pi ny)\bigr)^2_{L_2(\Gamma)}}{\bigl(v,v\bigr)_{L_2(\Gamma)}} \right\} 
\xrightarrow[\diam\{\Gamma\}\to 0]{} + \infty .
\end{equation}
\end{lemma}
{\bf Proof.}  The minimax principle implies that $I(\Gamma) \geq
I(\Gamma')$ for any $\Gamma'$ such that $\Gamma \subset \Gamma'$.
When $h = \diam\{\Gamma\}$ is small, one can choose $\Gamma' =
(p/q,(p+1)/q)$, with two integers $p$ and $q$.  For instance, one can
set $q = [1/h]$, where $[1/h]$ is the integer part of $1/h$ (the
largest integer that is less than or equal to $1/h$).  Since the
removal of a subsequence of positive terms does not increase the sum,
\begin{equation}
\sum\limits_{n=1}^\infty n \bigl(v , \sin(\pi ny)\bigr)^2_{L_2(\Gamma')} \geq 
\sum\limits_{k=1}^\infty n_k \bigl(v , \sin(\pi n_k y)\bigr)^2_{L_2(\Gamma')} ,
\end{equation}
one gets by setting $n_k = kq$:
\begin{equation}
\frac{\sum\limits_{n=1}^\infty n \bigl(v , \sin(\pi ny)\bigr)^2_{L_2(\Gamma')}}{\bigl(v,v\bigr)_{L_2(\Gamma')}} \geq 
q \frac{\sum\limits_{k=1}^\infty k \bigl(v , \sin(\pi k q y)\bigr)^2_{L_2(\Gamma')}}{\bigl(v,v\bigr)_{L_2(\Gamma')}} \, .
\end{equation}
Since the sine functions $\sin(\pi kq y)$ form a complete basis of
$L_2(\Gamma')$, the infimum of the right-hand side can be easily
computed and is equal to $q$, from which
\begin{equation}
I(\Gamma) \geq I(\Gamma') \geq q = [1/h] \geq \frac{1}{2h} \xrightarrow[h\to 0]{} + \infty 
\end{equation}
that completes the proof.  \qed

\begin{lemma}  \label{lem:1BD_continuity}
$\eta_1(\lambda)$ from (\ref{eq:mu1_1BD}) is a continuous function of
$\lambda$ for $\lambda \in (\nu_1, \nu_1 + \pi^2/a_1^2)$.
\end{lemma} \\
{\bf Proof.}  Let us denote the coefficients in (\ref{eq:Fv_1BD}) as
\begin{equation}
\beta_n(\lambda) = \begin{cases}  |\gamma_1| \bigl[\ctan(|\gamma_1| a_1) + \ctan(|\gamma_1| a_2)\bigr]  \quad (n=1), \cr
\gamma_n \bigl[\coth(\gamma_n a_1) + \coth(\gamma_n a_2)\bigr] \hskip 9mm (n \geq 2).  \end{cases}
\end{equation}
We recall that $|\gamma_1| = \sqrt{\lambda - \nu_1}$ and $\gamma_n =
\sqrt{\nu_n - \lambda}$.  Under the assumptions (\ref{eq:1BD_assump1},
\ref{eq:1BD_assump2}), one can easily check that all
$\beta_n(\lambda)$ are continuous functions of $\lambda$ when $\lambda
\in (\nu_1, \nu_1 + \pi^2/a_1^2)$.  In addition, all
$\beta_n(\lambda)$ with $n \geq 2$ have an upper bound uniformly on
$\lambda$
\begin{equation}
\beta_n(\lambda) \leq \sqrt{\nu_n - \lambda} ~ \bigl(\coth(\gamma_2 a_1) + \coth(\gamma_2 a_2)\bigr) \leq C \sqrt{\nu_n} ,
\end{equation}
where
\begin{equation}
C = \coth(\sqrt{\nu_2 - \nu_1}\, a_1) + \coth(\sqrt{\nu_2 - \nu_1}\, a_2)
\end{equation}
(here we replaced $\lambda$ by its minimal value $\nu_2$).  Under
these conditions, it was shown in \cite{Delitsyn04} that
$\eta_1(\lambda)$ is a continuous function of $\lambda$ that completes
the proof. \qed

\subsection{First eigenfunction}  \label{sec:1BD_first}

Our first goal is to show that the first eigenfunction $u$ (with the
smallest eigenvalue $\lambda$) is ``localized'' in the larger
subdomain when the opening $\Gamma$ is small enough.  By localization
we understand that the $L_2$-norm of the eigenfunction in the smaller
domain vanishes as the opening shrinks, i.e., $\diam\{\Gamma\}\to 0$.

We will obtain a stronger result by estimating the ratio of
$L_2$-norms of the eigenfunction in two arbitrary cross-sections
$\S_{x_1}$ and $\S_{x_2}$ on two sides of the barrier (i.e., with $x_1
\in (-a_1,0)$ and $x_2\in (0,a_2)$) and showing its divergence as the
opening shrinks.  We also discuss the rate of divergence.

Now we formulate the main
\begin{theorem}
Under assumptions (\ref{eq:1BD_assump1}, \ref{eq:1BD_assump2}), the
ratio of squared $L_2$-norms of the Dirichlet Laplacian eigenfunction
$u$ in two arbitrary cross-sections $\S_{x_1}$ (for any $x_1 \in
(-a_1,0)$) and $\S_{x_2}$ (for any $x_2\in (0,a_2)$) on two sides of
the barrier diverges as the opening $\Gamma$ shrinks, i.e.,
\begin{equation}
\frac{I(x_1)}{I(x_2)} \geq C \frac{\sin^2(|\gamma_1|(a_1+x_1))}{\sin(|\gamma_1| a_1)}  \xrightarrow[\diam\{\Gamma\}\to 0]{} +\infty  ,
\end{equation}
where $C > 0$ is a constant, and $I(x)$ is defined in
(\ref{eq:I_def}).
\end{theorem}  \\
{\bf Proof}.  According to Lemma \ref{lem:eigen_1BD}, $\lambda \to
\nu_1 + \pi^2/a_1^2$ as $\diam\{\Gamma\}\to 0$.  We denote then 
\begin{equation}
\lambda = \nu_1 + \pi^2/a_1^2 - \ve, 
\end{equation}
with a small parameter $\ve$ that vanishes as $\diam\{\Gamma\}\to 0$.
As a consequence, $\gamma_1$ is purely imaginary, and $|\gamma_1| a_1
\simeq \pi - a_1^2 \ve/(2\pi)$.  In turn, the assumption
(\ref{eq:1BD_assump2}) implies that all $\gamma_n$ with $n =
2,3,\ldots$ are positive.

We get then
\begin{eqnarray}  \nonumber
I(x_1) & = &  b_1^2  \frac{\sin^2(|\gamma_1|(a_1+x_1))}{\sin^2(|\gamma_1| a_1)}  
+ \sum\limits_{n=2}^\infty b_n^2 \frac{\sinh^2(\gamma_n(a_1+x_1))}{\sinh^2(\gamma_n a_1)}  \\
& \geq & b_1^2 \frac{\sin^2(|\gamma_1|(a_1+x_1))}{\sin^2(|\gamma_1| a_1)} , 
\end{eqnarray}
whereas
\begin{eqnarray} \nonumber
I(x_2) & = & b_1^2 \frac{\sin^2(|\gamma_1|(a_2-x_2))}{\sin^2(|\gamma_1| a_2)}  
+ \sum\limits_{n=2}^\infty b_n^2 \frac{\sinh^2(\gamma_n(a_2-x_2))}{\sinh^2(\gamma_n a_2)}  \\
& \leq & b_1^2 \frac{\sin^2(|\gamma_1|(a_2-x_2))}{\sin^2(|\gamma_1| a_2)} + \sum\limits_{n=2}^\infty b_n^2 ,  
\end{eqnarray}
because $\sinh(x)$ monotonously grows.  The last sum can be estimated
by rewriting the dispersion relation (\ref{eq:sum}):
\begin{eqnarray}  \nonumber  \hspace*{-4mm}
- b_1^2 |\gamma_1| \bigl[\ctan(|\gamma_1| a_1) + \ctan(|\gamma_1| a_2)\bigr] & = & 
\sum\limits_{n=2}^\infty b_n^2 \gamma_n \bigl[\coth(\gamma_n a_1) + \coth(\gamma_n a_2)\bigr]  \\
& \geq & \gamma_2 \bigl[\coth(\gamma_2 a_1) + \coth(\gamma_2 a_2)\bigr] \sum\limits_{n=2}^\infty b_n^2 , 
\end{eqnarray}
because $x \coth(x)$ monotonously grows.  As a consequence, one gets
\begin{equation}
\begin{split}
I(x_2) & \leq b_1^2 \left[ \frac{1}{\sin^2(|\gamma_1| a_2)}  
- \frac{|\gamma_1| \bigl[\ctan(|\gamma_1| a_1) + \ctan(|\gamma_1| a_2)\bigr]}{\gamma_2 \bigl[\coth(\gamma_2 a_1) + \coth(\gamma_2 a_2)\bigr]} \right], \\
\end{split}
\end{equation}
where we used $\sin^2(|\gamma_1|(a_2-x_2)) \leq 1$.  We conclude that
\begin{equation}
\label{eq:1BD_ratio}
\frac{I(x_1)}{I(x_2)} \geq C_\lambda \frac{\sin^2(|\gamma_1|(a_1+x_1))}{\sin(|\gamma_1| a_1)} ,
\end{equation}
where
\begin{equation}
\label{eq:1BD_C}
C_\lambda = \left[ \frac{\sin(|\gamma_1| a_1)}{\sin^2(|\gamma_1| a_2)}  
- \frac{|\gamma_1| \bigl[\cos(|\gamma_1| a_1) + \sin(|\gamma_1| a_1)\ctan(|\gamma_1| a_2)\bigr]}
{\gamma_2 \bigl[\coth(\gamma_2 a_1) + \coth(\gamma_2 a_2)\bigr]} \right]^{-1} .
\end{equation}
Since $|\gamma_1| a_1 \to \pi$ as $\diam\{\Gamma\} \to 0$, the
denominator $\sin(|\gamma_1| a_1)$ in (\ref{eq:1BD_ratio}) diverges,
while $C_\lambda$ in (\ref{eq:1BD_C}) converges to a strictly positive
constant (because $a_2 < a_1$) that completes the proof of the
theorem. \qed

\begin{remark}
If two subdomains $\Omega_1$ and $\Omega_2$ are equal, i.e., $a_1 =
a_2$, the first eigenfunction cannot be localized due to the
reflection symmetry: $u(x,\y) = u(-x,\y)$.  While the estimate
(\ref{eq:1BD_ratio}) on the ratio of two squared $L_2$ norms remains
valid, the constant $C_\lambda$ in (\ref{eq:1BD_C}) vanishes as
$|\gamma_1|a_1 \to \pi$ because $\sin(|\gamma_1|a_2) \to 0$.
\end{remark}

\begin{remark}
The assumption (\ref{eq:1BD_assump2}) was used to ensure that
$\gamma_n$ with $n \geq 2$ were positive while the corresponding modes
were exponentially decaying.  We checked numerically that this
assumption is not a necessary condition for localization (see Appendix
\ref{sec:numerics}).
\end{remark}

\begin{remark}
The above formulation of the mode matching method can be extended to
two cylinders of different cross-sections $\S^1$ and $\S^2$ connected
through an opening set $\Gamma$: $\Omega = \bigl(([-a_1,0]\times
\S^1) \cup ([0,a_2]\times \S^2)\bigr) \backslash \Gamma$.  For
instance, the eigenfunction representation (\ref{eq:1BD_u12}) would
read as
\begin{eqnarray}
\label{eq:1BD_u12gen}
u_1(x,\y) &=& \sum\limits_{n=1}^\infty \bigl(u_{|\Gamma}, \psi^1_n\bigr)_{L_2(\Gamma)} ~ \psi^1_n(\y) ~ 
\frac{\sinh(\gamma^1_n(a_1+x))}{\sinh(\gamma^1_n a_1)}  \quad (-a_1 < x < 0), \\
u_2(x,\y) &=& \sum\limits_{n=1}^\infty \bigl(u_{|\Gamma}, \psi^2_n\bigr)_{L_2(\Gamma)} ~ \psi^2_n(\y) ~ 
\frac{\sinh(\gamma^2_n(a_2-x))}{\sinh(\gamma^2_n a_2)}  \quad (0 < x < a_2), 
\end{eqnarray}
with $\gamma_n^{1,2} = \sqrt{\nu_n^{1,2} - \lambda}$, where
$\nu_n^{1,2}$ and $\psi_n^{1,2}$ are the eigenvalues and
eigenfunctions of $\Delta_\perp$ in cross-sections $\S^1$ and $\S^2$.
For instance, the dispersion relation reads
\begin{equation}
\sum\limits_{n=1}^\infty \biggl[\bigl(u_{|\Gamma}, \psi^1_n\bigr)^2_{L_2(\Gamma)}  \gamma_n^1 \coth(\gamma^1_n a_1)
+ \bigl(u_{|\Gamma}, \psi^2_n\bigr)^2_{L_2(\Gamma)} \gamma_n^2 \coth(\gamma^2_n a_2) \biggr] = 0 .
\end{equation}
The remaining analysis would be similar although statements about
localization would be more subtle.  In particular, the localization of
the first eigenfunction does not necessarily occur in the subdomain
with larger Lebesgue measure \cite{Delitsyn12b}. 
\end{remark}

\subsection{Higher-order eigenfunctions}   \label{sec:1BD_other}

Similar arguments can be applied to show localization of other
eigenfunctions.  When the eigenvalue $\lambda$ is progressively
increased, there are more and more purely imaginary $\gamma_n$ and
thus more and more oscillating terms in the representation of an
eigenfunction.  These oscillations start to interfere with each other,
and localization is progressively reduced.  When the characteristic
wavelength $1/\sqrt{\lambda}$ becomes comparable to the size of the
opening, no localization is expected.  From these qualitative
arguments, it is clear that proving localization for higher-order
eigenfunctions becomes more challenging while some additional
constraints are expected to appear.  To illustrate these difficulties,
we consider an eigenfunction of the Dirichlet Laplacian for which
$|\gamma_2| a_1 \to \pi$, i.e.,
\begin{equation}
\lambda = \nu_2 + \pi^2/a_1^2 + \ve, 
\end{equation}
with $\ve \to 0$.  

As earlier for the first eigenfunction, we estimate the squared $L_2$
norm of this eigenfunction in two arbitrary cross-sections $\S_{x_1}$
and $\S_{x_2}$.  From (\ref{eq:I_1BD}), we have
\begin{equation}
I(x_1) \leq  b_2^2 \frac{\sin^2(|\gamma_2| (a_1+x_1))}{\sin^2(|\gamma_2|a_1)} ,
\end{equation}
where we kept only the leading term with $n=2$ (for which the
denominator diverges).

In order to estimate $I(x_2)$, we use the dispersion relation
(\ref{eq:sum}) to get
\begin{eqnarray} \nonumber
&& - b_2^2 |\gamma_2|\bigl[\ctan(|\gamma_2|a_1) + \ctan(|\gamma_2|a_2)\bigr] \\  \nonumber
&=& b_1^2 |\gamma_1|\bigl[\ctan(|\gamma_1|a_1) + \ctan(|\gamma_1|a_2)\bigr] 
+ \sum\limits_{n=3}^\infty b_n^2 \gamma_n \bigl[\coth(\gamma_n a_1) + \coth(\gamma_n a_2) \bigr]  \\  \nonumber
&\geq& b_1^2 |\gamma_1|\bigl[\ctan(|\gamma_1|a_1) + \ctan(|\gamma_1|a_2)\bigr] 
+ \gamma_3 \bigl[\coth(\gamma_3 a_1) + \coth(\gamma_3 a_2) \bigr]  \sum\limits_{n=3}^\infty b_n^2  \\
\label{eq:1BD_auxil4}
&\geq& C \biggl[b_1^2 + \sum\limits_{n=3}^\infty b_n^2 \biggr],  
\end{eqnarray}
where 
\begin{equation}
C = \min \biggl\{ |\gamma_1|\bigl[\ctan(|\gamma_1|a_1) + \ctan(|\gamma_1|a_2)\bigr] , \gamma_3 \bigl[\coth(\gamma_3 a_1) + \coth(\gamma_3 a_2) \bigr]\biggr \} > 0  ,
\end{equation}
and we assumed that
\begin{equation}
\label{eq:1BD_auxil5}
\ctan(|\gamma_1|a_1) + \ctan(|\gamma_1|a_2) > 0 .
\end{equation}

Using (\ref{eq:1BD_auxil4}), we get
\begin{eqnarray} \nonumber
I(x_2) &=& b_1^2 \frac{\sin^2(|\gamma_1|(a_2-x_2))}{\sin^2(|\gamma_1|a_2)} + b_2^2 \frac{\sin^2(|\gamma_2|(a_2-x_2))}{\sin^2(|\gamma_2|a_2)} + 
\sum\limits_{n=3}^\infty b_n^2 \frac{\sinh^2(\gamma_n (a_2-x_2))}{\sinh^2(\gamma_n a_2)}  \\  \nonumber
& \leq&  \frac{b_1^2}{\sin^2(|\gamma_1|a_2)} +  \frac{b_2^2}{\sin^2(|\gamma_2|a_2)} + \sum\limits_{n=3}^\infty b_n^2  \\  \nonumber
& \leq&  \frac{b_2^2}{\sin^2(|\gamma_2|a_2)} + \frac{1}{\sin^2(|\gamma_1|a_2)} \biggl[b_1^2 + \sum\limits_{n=3}^\infty b_n^2 \biggr]  \\  
& \leq& b_2^2 \biggl[\frac{1}{\sin^2(|\gamma_2|a_2)} - 
\frac{|\gamma_2|\bigl[\ctan(|\gamma_2|a_1) + \ctan(|\gamma_2|a_2)\bigr]}{C \sin^2(|\gamma_1|a_2)}\biggr] . 
\end{eqnarray}
We finally obtain
\begin{equation}
\label{eq:nBD_ratio}
\frac{I(x_1)}{I(x_2)} \geq C_\lambda \frac{\sin^2(|\gamma_2| (a_1+x_1))}{\sin(|\gamma_2|a_1)}  ,
\end{equation}
with
\begin{equation}
C_\lambda = \left(\frac{\sin(|\gamma_2|a_1)}{\sin^2(|\gamma_2|a_2)} 
- \frac{|\gamma_2|\bigl[\cos(|\gamma_2|a_1) + \sin(|\gamma_2|a_1) \ctan(|\gamma_2|a_2)\bigr]}{C \sin^2(|\gamma_1|a_2)}\right)^{-1} .
\end{equation}
When $|\gamma_2|a_1 \to \pi$, the denominator in (\ref{eq:nBD_ratio})
diverges while $C_\lambda$ converges to a strictly positive constant
if $\sin(|\gamma_1|a_2)$ does not vanish.  This additional constraint
can be formulated as
\begin{equation}
\label{eq:1BD_auxil6}
\sqrt{a_2^2(\nu_2 - \nu_1)/\pi^2 + a_2^2/a_1^2} \notin \N.
\end{equation}
%This completes the proof. \qed

\begin{remark}
The additional constraints (\ref{eq:1BD_auxil5}, \ref{eq:1BD_auxil6})
were used to ensure that $|\gamma_1|a_2$ does not converge to a
multiple of $\pi$ as $|\gamma_2|a_1 \to \pi$.  Given that the lengths
$a_1$ and $a_2$ can in general be chosen arbitrarily, one can easily
construct such domains, in which this condition is not satisfied.  For
instance, setting $|\gamma_1|a_2 = |\gamma_2|a_1 = \pi$, one gets the
relation $1/a_2^2 - 1/a_1^2 = (\nu_2 - \nu_1)/\pi^2$ under which the
constraint is not fulfilled.  At the same time, numerical evidence
(see Fig. \ref{fig:eigen_1BD2}) suggests that the constraint
(\ref{eq:1BD_auxil6}) may potentially be relaxed.
\end{remark}

\begin{remark}
One can also consider eigenfunctions for which $|\gamma_n|a_1 \to
\pi$.  These modes exhibit ``one oscillation'' in the lateral
direction $x$ and ``multiple oscillations'' in the transverse
directions $\y$.  The analysis is very similar but additional
constrains may appear.  In turn, the analysis would be more involved
in a more general situation when $|\gamma_n|a_1 \to \pi k$, with an
integer $k$.  Finally, one can investigate the eigenfunctions
localized in the smaller domain $\Omega_2$.  In this situation, which
is technically more subtle, one can get similar estimates on the ratio
of squared $L_2$ norms.  Moreover, Theorems \ref{lem:main_1BD} and
\ref{lem:eigen_1BD} ensure the existence of an eigenvalue $\lambda$,
which is close to the first eigenvalue in the smaller domain, $\lambda
\to \nu_1 + \pi^2/a_2^2$, implying $|\gamma_1|a_2 \to \pi$.  We do not
provide rigorous statements for these extensions but some numerical
examples are given in Appendix \ref{sec:numerics}.
\end{remark}

%%%%%%%%%%%%  TWO BARRIERS  %%%%%%%%%%%%%%%%%%
\section{Scattering problem with two barriers}
\label{sec:2BD}

\begin{figure} 
\begin{center}
\includegraphics[width=120mm]{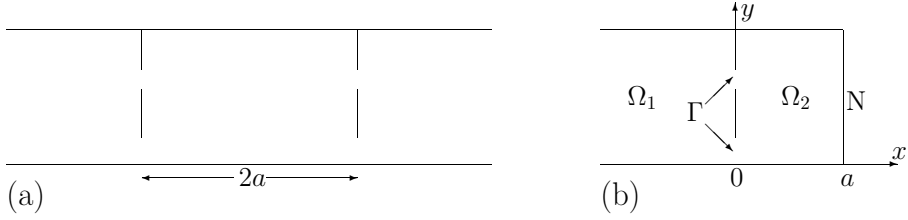} % {domain_twobarriers2.eps} 
\end{center} 
\caption{
{\bf (a)} An infinite cylinder with two identical barriers at distance
$2a$.  {\bf (b)} The half of the above domain, i.e., a semi-infinite
cylinder with a single barrier at $x = 0$ and Neumann boundary
condition at $x = a$.  Although this schematic illustration is shown
in two dimensions, the results are valid for a general cylinder $\R
\times \S$ of arbitrary bounded cross-section $\S\subset \R^n$ with a
piecewise smooth boundary $\partial\S$ and arbitrary opening
$\Gamma\subset \S$.}
\label{fig:domain_TB} 
\end{figure}

We consider a scattering problem for an infinite cylinder $\tilde
\Omega_0 = \R \times \S$ of arbitrary bounded cross-section $\S\subset
\R^d$ with a piecewise smooth boundary $\partial\S$, with two barriers
located at $x = 0$ and $x = 2a$.  In other words, for a given
``opening'' $\Gamma \subset \S$ inside the cross-section $\S$, we
consider the domain $\tilde \Omega = \tilde \Omega_0
\backslash (\{0,2a\} \times (\S\backslash \Gamma))$
(Fig. \ref{fig:domain_TB}).  Due to the reflection symmetry at $x =
a$, this domain can be replaced by another domain, $\Omega_0 =
(-\infty,a)\times \S$, with a single barrier at $x = 0$ so that
$\Omega = \Omega_0 \backslash (\{0\} \times (\S\backslash \Gamma))$.
This barrier splits the domain $\Omega$ into two subdomains:
$\Omega_1$ (for $x < 0$) and $\Omega_2$ (for $0 < x < a$).  As
earlier, we denote the cross-section at $x$ as $\S_x = \{x\}
\times \S$.

An acoustic wave $u$ with the wave number $\sqrt{\lambda}$ satisfies
the following equation
\begin{equation}
- \Delta u = \lambda u \quad \mbox{in~} \Omega,  \quad u_{|\pa\backslash \S_a} = 0, \quad  \left. \frac{\partial u}{\partial x}\right|_{\S_a} = 0,
\end{equation}
i.e., with Dirichlet boundary condition everywhere on $\pa$ except for
the cross-section $\S_a$ at which Neumann boundary condition is
imposed to respect the reflection symmetry.  In contrast to spectral
problems considered in previous sections, the squared wave number,
$\lambda$, is fixed, and one studies the propagation of such a wave
along the waveguide.

As earlier, we consider the auxiliary Dirichlet eigenvalue problem
(\ref{eq:eigen_perp_1BD}) in the cross-section $\S$ with ordered
eigenvalues $\nu_k$, and set $\gamma_n = \sqrt{\nu_n - \lambda}$.  The
first eigenvalue $\nu_1$ determines the cut-off frequency below which
no wave can travel in the waveguide of the cross-section $\S$.  Here
we focus on the waves near cut-off frequency and we assume that
\begin{equation}
\label{eq:2B_assump0}
\nu_1 < \lambda < \nu_2 .
\end{equation}
As a consequence, $\gamma_1$ is purely imaginary while all other
$\gamma_n$ are positive.  

The wave $u$ in the subdomain $\Omega_1$ can be written as
\begin{equation}
\label{eq:u_TB}
u_1(x,\y) = e^{i|\gamma_1|x} \psi_1 + c_1 e^{-i|\gamma_1|x} \psi_1 + \sum\limits_{n=2}^\infty c_n e^{\gamma_n x} \psi_n  ,
\end{equation}
where the coefficients $c_n$ will be determined below.  The first term
represents the coming wave, while the other terms represent reflected
waves.  Note that the terms with $n\geq 2$ are exponentially vanishing
for $x< 0$.

\begin{theorem}
For any opening $\Gamma$ with a small enough Lebesgue measure
$|\Gamma|$, there exists the critical wave number $\sqrt{\lambda_c}$
at which the wave is fully propagating across two barriers, i.e., $c_1
= 1$.
\end{theorem} \\
{\bf Proof.}  The proof consists in two steps: (i) we establish an
explicit equation that determines $c_1$ for any opening $\Gamma$, and
(ii) we prove the existence of its solution $\lambda_c$ when
$|\Gamma|$ is small enough.

{\bf Step 1}.  Taking the scalar product of (\ref{eq:u_TB}) at $x = 0$
with $\psi_1$, one finds
\begin{equation}
\label{eq:c1_TB}
c_1 = (u_{|\Gamma}, \psi_1)_{L_2(\Gamma)} - 1 .
\end{equation}
When the scalar product is $0$, $c_1 = -1$ so that one gets a standing
wave in the region $x < 0$.  In turn, if the scalar product is $2$,
one gets $c_1 = 1$ that corresponds to a fully propagating wave.  In
what follows, we will investigate these cases.

Similarly, one finds $c_n = (u_{|\Gamma}, \psi_n)_{L_2(\Gamma)}$ so
that the wave in the first domain $\Omega_1$ is
\begin{equation}
u_1(x,\y) = e^{i|\gamma_1|x} \psi_1 + e^{-i|\gamma_1|x} \biggl[\bigl(u_{|\Gamma}, \psi_1\bigr)_{L_2(\Gamma)} - 1\biggr]  \psi_1 
+ \sum\limits_{n=2}^\infty e^{\gamma_n x} \bigl(u_{|\Gamma}, \psi_n\bigr)_{L_2(\Gamma)}  \psi_n  .
\end{equation}
In the second domain $\Omega_2$, the wave reads as
\begin{eqnarray} \nonumber
u_2(x,\y) &=& \frac{\cos(|\gamma_1|(a-x))}{\cos(|\gamma_1| a)} \bigl(u_{|\Gamma}, \psi_1\bigr)_{L_2(\Gamma)} \psi_1\\
&+& \sum\limits_{n=1}^\infty \frac{\cosh(\gamma_n (a-x))}{\cosh(\gamma_n a)} \bigl(u_{|\Gamma}, \psi_n\bigr)_{L_2(\Gamma)}  \psi_n  \qquad (0 < x < a),
\end{eqnarray}
where we separated the first oscillating term from the remaining
exponentially decaying terms.  Matching the derivatives $u'_1$ and
$u'_2$ with respect to $x$ at the opening $\Gamma$, one gets for any
$\y \in \Gamma$:
\begin{eqnarray*}
&& 2i|\gamma_1| \psi_1(\y) - i|\gamma_1| \bigl(u_{|\Gamma}, \psi_1\bigr)_{L_2(\Gamma)} \psi_1(\y) + 
\sum\limits_{n=2}^\infty \gamma_n \bigl(u_{|\Gamma}, \psi_n\bigr)_{L_2(\Gamma)}  \psi_n(\y) \\
&=& |\gamma_1| \bigl(u_{|\Gamma}, \psi_1\bigr)_{L_2(\Gamma)} \tan(|\gamma_1| a)) \psi_1(\y)
- \sum\limits_{n=2}^\infty \gamma_n \bigl(u_{|\Gamma}, \psi_n\bigr)_{L_2(\Gamma)} \tanh(\gamma_n a) \psi_n(\y) ,
\end{eqnarray*}
or, in a shorter form,
\begin{equation}
A u_{|\Gamma} - \beta \bigl(u_{|\Gamma}, \psi_1\bigr)_{L_2(\Gamma)} \psi_1(\y) = -2i|\gamma_1| \psi_1(\y) \quad (\y \in \Gamma) ,
\end{equation}
where
\begin{equation}
\beta =  i |\gamma_1| + |\gamma_1|\tan(|\gamma_1| a) + 1 + \tanh(|\gamma_1| a) ,
\end{equation}
and we introduced a positive-definite self-adjoint operator $A$ acting
on a function $v$ from $H^{\frac12}(\Gamma)$ as
\begin{equation}
\label{eq:A_def}
A v = \sum\limits_{n=1}^\infty \beta_n \bigl(v, \psi_n\bigr)_{L_2(\Gamma)}  \psi_n(\y)  ,
\end{equation}
with 
\begin{equation}
\label{eq:betan}
\beta_n = \begin{cases} ~~~~ 1 + \tanh(|\gamma_1| a)  \qquad (n=1),  \cr 
\gamma_n \bigl(1 + \tanh(\gamma_n a)\bigr)  \qquad (n>1). \end{cases}
\end{equation}
Since the coefficients $\beta_n$ are strictly positive, the operator
$A$ can be inverted to get
\begin{equation}
u_{|\Gamma} - \beta \bigl(u_{|\Gamma}, \psi_1\bigr)_{L_2(\Gamma)} A^{-1} \psi_1  = -2i|\gamma_1| A^{-1} \psi_1  \quad (\y \in \Gamma).
\end{equation}
Multiplying this relation by $\psi_1$ and integrating over $\Gamma$,
one finds
\begin{equation}
\bigl(u_{|\Gamma}, \psi_1\bigr)_{L_2(\Gamma)} = - \frac{2i|\gamma_1| \bigl(A^{-1} \psi_1, \psi_1\bigr)_{L_2(\Gamma)}}
{1 - \beta \bigl(A^{-1} \psi_1, \psi_1\bigr)_{L_2(\Gamma)}}  = \frac{2i|\gamma_1|}{i|\gamma_1| + \eta(\lambda)}  ,  
\end{equation}
where
\begin{equation}
\label{eq:2B_eta}
\eta(\lambda) \equiv |\gamma_1| \tan(|\gamma_1| a) + 1 + \tanh(|\gamma_1| a) - \frac{1}{\bigl(A^{-1} \psi_1, \psi_1\bigr)_{L_2(\Gamma)}} .
\end{equation}

If there exists $\lambda_c$ such that $\eta(\lambda_c) = 0$, then
$\bigl(u_{|\Gamma}, \psi_1\bigr)_{L_2(\Gamma)} = 2$ at this
$\lambda_c$ and thus $c_1 = 1$ from (\ref{eq:c1_TB}) that would
complete the proof.

{\bf Step 2}.  It is easy to show that $\eta(\lambda)$ is a continuous
function of $\lambda$ for $\lambda \in (\nu_1, \nu_1 + \pi^2/(4a^2))$.
In fact, all $|\gamma_n|$ are continuous for any $\lambda$, $\tan(x)$
is continuous on the interval $(0,\pi/2)$ and thus for $x =
|\gamma_1|a = \sqrt{\lambda-\nu_1}\, a$.  Finally, the continuity of
$\bigl(A^{-1} \psi_1, \psi_1\bigr)_{L_2(\Gamma)}$ for any $\lambda$
was shown in \cite{Delitsyn04}.  In what follows, we re-enforce the
assumption (\ref{eq:2B_assump0}) as
\begin{equation}
\lambda \in \Lambda,  \qquad \Lambda = (\nu_1,\, \min\{ \nu_2,\, \nu_1 + \pi^2/(4a^2) \}) .
\end{equation}

According to Lemma \ref{lem:Apsi}, for any fixed $\lambda \in
\Lambda$, the scalar product $\bigl(A^{-1} \psi_1,
\psi_1\bigr)_{L_2(\Gamma)}$ can be made arbitrarily small by taking
the opening $\Gamma$ small enough.  As a consequence, if the opening
$\Gamma$ is small enough, there exists $\lambda$ such that
$\eta(\lambda) < 0$.

Now, fixing $\Gamma$, we vary $\lambda$ in such a way that $|\gamma_1|
a \to \pi/2$.  Since $\tan(|\gamma_1| a)$ grows up to infinity, the
first term in (\ref{eq:2B_eta}) becomes dominating, and
$\eta(\lambda)$ gets positive values.  We conclude thus that there
exists $\lambda_c$ at which $\eta(\lambda_c) = 0$.  This completes the
proof. \qed

\begin{lemma}
\label{lem:Apsi}
For any $\lambda\in\Lambda$ and any $\ve > 0$, there exists $\delta >
0$ such that for $|\Gamma| < \delta$, one has $\bigl(A^{-1} \psi_1,
\psi_1\bigr)_{L_2(\Gamma)} < \ve$.  
\end{lemma} \\
{\bf Proof.}  Denoting $\phi = A^{-1} \psi_1$, we can write $A \phi =
\psi_1$ as
\begin{equation}
\sum\limits_{n=1}^{\infty} \beta_n \bigl(\phi, \psi_n\bigr)_{L_2(\Gamma)} ~ \psi_n = \psi_1  \quad (\y \in \Gamma),
\end{equation}
with $\beta_n$ given by (\ref{eq:betan}).
Multiplying this relation by $\phi$ and integrating over $\Gamma$
yield
\begin{equation}
\sum\limits_{n=1}^{\infty} \beta_n \bigl(\phi, \psi_n\bigr)^2_{L_2(\Gamma)} = \bigl(\phi, \psi_1\bigr)_{L_2(\Gamma)} .
\end{equation}
Since the coefficients $\beta_n$ asymptotically grow with $n$, one
gets the following estimate
\begin{equation}
\bigl(\phi, \psi_1\bigr)_{L_2(\Gamma)} \geq C_\lambda \sum\limits_{n=1}^{\infty} \bigl(\phi, \psi_n\bigr)^2_{L_2(\Gamma)} = C_\lambda ||\phi||^2_{L_2(\Gamma)} ,
\end{equation}
where
\begin{equation}
C_\lambda = \min\limits_{n\geq 1} \{\beta_n \}  > 0 .
\end{equation}
On the other hand, $\bigl(\phi, \psi_1\bigr)_{L_2(\Gamma)} \leq
||\phi||_{L_2(\Gamma)} ~ ||\psi_1||_{L_2(\Gamma)}$ so that
\begin{equation}
C_\lambda \, ||\phi||_{L_2(\Gamma)} \leq  ||\psi_1||_{L_2(\Gamma)} .
\end{equation}
Note that
\begin{equation}
||\psi_1||^2_{L_2(\Gamma)} = \int\limits_\Gamma d\y ~ |\psi_1(\y)|^2 \leq |\Gamma| ~ \max\limits_{\y\in \Gamma} \{ |\psi_1(\y)|^2 \} 
\leq |\Gamma| ~ \max\limits_{\y\in \S} \{ |\psi_1(\y)|^2 \} .
\end{equation}
Since the maximum of an eigenfunction $\psi_1$ is fixed by its
normalization in the cross-section $\S$ (and does not depend on
$\Gamma$), we conclude that $||\psi_1||_{L_2(\Gamma)}$ vanishes as the
opening $\Gamma$ shrinks.  Finally, we have
\begin{equation}
0 \leq \bigl(A^{-1} \psi_1, \psi_1\bigr)_{L_2(\Gamma)} \leq ||A^{-1} \psi_1||_{L_2(\Gamma)} ~ ||\psi_1||_{L_2(\Gamma)} 
\leq \frac{1}{C_\lambda} ||\psi_1||^2_{L_2(\Gamma)}  \xrightarrow[|\Gamma| \to 0]{} 0  
\end{equation}
that completes the proof of the lemma. \qed

%%%%%%%%%%%  CIPOLLINO  %%%%%%%%%%%%%%%%%
\section{Geometry-induced localization in a noncylindrical domain}
\label{sec:cipollino}

\subsection{Preliminaries}

Examples from previous sections relied on the orthogonality of the
lateral coordinate $x$ and the transverse coordinates $\y$.  In this
section, we illustrate an application of mode matching methods to
another situation admitting the separation of variables.

\begin{figure}
\begin{center}
\includegraphics[width=60mm]{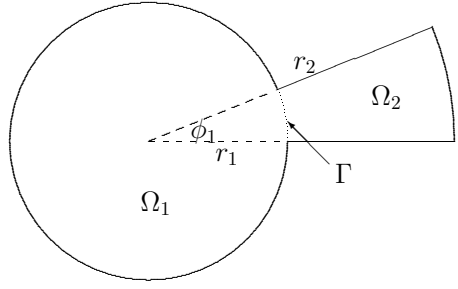} % domain_cippolino.eps}
\end{center}
\caption{
The domain $\Omega$ is decomposed into a disk $\Omega_1$ of radius
$R_1$, and a part of a circular sector $\Omega_2$ of angle $\phi_1$
between two circles of radii $R_1$ and $R_2$.  Two domains are
connected through an opening $\Gamma$ (an arc $(0,\phi_1)$ on the
circle of radius $R_1$).}
\label{fig:domain3}
\end{figure}

We consider the planar domain
\begin{equation}
\Omega = \oe \Omega_1 \cup \oe \Omega_2 ,
\end{equation}
where $\Omega_1$ is the disk of radius $R_1$, $\Omega_1 = \left\{ 0 <
r < R_1, 0 \leq \phi \leq 2 \pi \right\}$, and $\Omega_2$ is a part of
a circular sector of angle $\phi_1$ between two circles of radii $R_1$
and $R_2$: $\Omega_2 = \left\{ R_1 < r < R_2, 0 < \phi < \phi_1
\right\}$ (see Fig. \ref{fig:domain3}).  In general, one can consider
a disk with several sector-like ``petals''.  We focus on the Dirichlet
eigenvalue problem
\begin{equation}
\label{eq:eigen_disk}
- \Delta u = \lambda u,    \qquad u|_{\dc \Omega} = 0 .
\end{equation}

In polar coordinates ($r,\phi$), the separation of variables allows
one to write explicit representations $u_1(r,\phi)$ and $u_2(r,\phi)$
of the solution of (\ref{eq:eigen_disk}) in domains $\Omega_1$ and
$\Omega_2$ as
\begin{eqnarray} 
u_1(r,\phi) &=& \frac{1}{2 \pi} \frac{J_0(\sqrt{\lambda}\, r)}{J_0(\sqrt{\lambda}\, R_1)} \bigl(u_{|\Gamma}, 1\bigr)_{L_2(\Gamma)} \\
\nonumber
&+& \frac{1}{\pi} \sum \limits_{n=1}^{\infty}
\frac{J_n(\sqrt{\lambda}\, r)}{J_n(\sqrt{\lambda}\, R_1)} \biggl( \bigl(u_{\Gamma}, \cos n \phi)_{L_2(\Gamma)} \cos n \phi +
\bigl(u_{|\Gamma}, \sin n \phi \bigr)_{L_2(\Gamma)} \sin n \phi \biggr)
\end{eqnarray}
and
\begin{equation} 
u_2(r,\phi) = \frac{2}{\phi_1} \sum \limits_{n=1}^{\infty} \frac{\psi_n(\sqrt{\lambda}\, r)}{\psi_n(\sqrt{\lambda}\, R_1)} 
\bigl(u_{\Gamma}, \sin \alpha_n \phi \bigr)_{L_2(\Gamma)} \sin \alpha_n \phi ,
\end{equation}
where 
\begin{equation}
\alpha_n = \frac{\pi}{\phi_1} n, 
\end{equation}
and $\psi_n(\sqrt{\lambda}\, r)$ are solutions of the Bessel equation
satisfying the Dirichlet boundary condition at $r = R_2$
($\psi_n(\sqrt{\lambda}\, R_2) = 0$):
\begin{equation}
\label{eq:psin}
\psi_n(r) = J_{\alpha_n}(r) \, Y_{\alpha_n}(\sqrt{\lambda}\, R_2) - Y_{\alpha_n}(r) \, J_{\alpha_n}(\sqrt{\lambda}\, R_2),
\end{equation}
and $J_n(z)$ and $Y_n(z)$ are Bessel functions of the first and second
kind.

Since the eigenfunction $u$ is analytic in $\Omega$, its radial
derivatives match at the opening $\Gamma$:
\begin{eqnarray} \nonumber
&& \frac{1}{2\pi} \frac{J'_0(\sqrt{\lambda}\, R_1)}{J_0(\sqrt{\lambda}\, R_1)} \bigl(u_{|\Gamma}, 1)_{L_2(\Gamma)} \\
\nonumber
&& + \frac{1}{\pi} \sum \limits_{n=1}^{\infty}
\frac{J'_n(\sqrt{\lambda}\, R_1)}{J_n(\sqrt{\lambda}\, R_1)} \biggl(\bigl(u_{|\Gamma}, \cos n \phi\bigr)_{L_2(\Gamma)} \cos n \phi +
\bigl(u_{|\Gamma}, \sin n \phi)_{L_2(\Gamma)} \sin n \phi \biggr) \\
&& = \frac{2}{\phi_1} \sum \limits_{n=1}^{\infty} \frac{\psi'_n(\sqrt{\lambda}\, R_1)}{\psi_n(\sqrt{\lambda}\, R_1)} 
\bigl(u_{|\Gamma}, \sin \alpha_n \phi\bigr)_{L_2(\Gamma)} \sin \alpha_n \phi  \qquad (0 < \phi < \phi_1).
\end{eqnarray}
Multiplying this equation by $u_{|\Gamma}$ and integrating over
$\Gamma$ yield the dispersion relation
\begin{eqnarray} \nonumber
&& \frac{1}{2 \pi} \frac{J'_0(\sqrt{\lambda}\, R_1)}{J_0(\sqrt{\lambda}\, R_1)} \bigl(u_{|\Gamma}, 1\bigr)^2_{L_2(\Gamma)} \\
\nonumber
&& + \frac{1}{\pi} \sum\limits_{n=1}^{\infty}
\frac{J'_n(\sqrt{\lambda}\, R_1)}{J_n(\sqrt{\lambda}\, R_1)} \biggl(\bigl( u_{|\Gamma}, \cos n \phi \bigr)^2_{L_2(\Gamma)} +
\bigl(u_{|\Gamma}, \sin n \phi\bigr)^2_{L_2(\Gamma)} \biggr)  \\
\label{eq:dispersion_disk}
&& = \frac{2}{\phi_1} \sum\limits_{n=1}^{\infty} \frac{\psi'_n(\sqrt{\lambda}\, R_1)}{\psi_n(\sqrt{\lambda}\, R_1)} 
\bigl(u_{|\Gamma}, \sin \alpha_n \phi \bigr)^2_{L_2(\Gamma)}  .
\end{eqnarray}

Our goal is to show that there exists an eigenfunction $u$ which is
localized in the ``petal'' $\Omega_2$ and negligible in the disk
$\Omega_1$.  For this purpose, we consider an auxiliary Dirichlet
eigenvalue problem in the sector $\Omega_3 = \{ 0 < r < R_2,~ 0 < \phi
< \phi_1\}$ for which all eigenvalues and eigenfunctions are known
explicitly.  These eigenfunctions are natural candidates to ``build''
localized eigenfunctions in $\Omega$.

We proceed as follows.  In Sec. \ref{sec:Cip_localization}, we show
the existence of an eigenvalue $\lambda$ of the Dirichlet Laplacian in
$\Omega$ which is close to the first eigenvalue $\mu$ of the Dirichlet
Laplacian in the sector $\Omega_3$.  In Sec. \ref{sec:Cip_estimates},
we estimate the $L_2$-norm of an eigenfunction in $\Omega$.  These
estimates rely on some technical inequalities on Bessel functions that
we prove in Appendix \ref{sec:Cip_inequalities}.  These steps reveal
restrictions on three geometric parameters of $\Omega$: two radii
$R_1$ and $R_2$, and the angle $\phi_1$.  We will show that
localization occurs for thin long ``petals'' (i.e., large $R_2$ and
small $\phi_1$).  The radius $R_1$ of the disk should be small as
compared to $R_2$.  In particular, we set
\begin{equation}
R_1 \leq \frac{j'_1}{\sqrt{\lambda}} ,
\end{equation}
where $j'_1$ is the first zero of $J'_1(z)$.

\subsection{Localization in $\Omega_2$}   \label{sec:Cip_localization}

In order to prove the existence of an eigenfunction $u$ in $\Omega$
which is localized in $\Omega_2$, we consider an auxiliary eigenvalue
problem for the Dirichlet Laplacian in the sector $\Omega_3 = \left\{0
< r < R_2,~ 0 < \phi < \phi_1 \right\}$.  For this domain, all
eigenvalues and eigenfunctions are known explicitly.  In particular,
the first eigenvalue $\mu$ and the corresponding eigenfunction are
\begin{equation}
\label{eq:Cip_mu}
\mu = \frac{j_{\alpha_1}^2}{R_2^2},  \qquad  v(r,\phi) = C_v \, J_{\alpha_1}(\sqrt{\mu}\, r) \, \sin(\pi \phi/\phi_1) ,
\end{equation}
where $\alpha_1 = \pi/\phi_1$, $j_{\alpha_1}$ is the first zero of the
Bessel function $J_{\alpha_1}(z)$, $J_{\alpha_1}(j_{\alpha_1}) = 0$,
and $C_v$ is the normalization constant to ensure
$||v||_{L_2(\Omega_3)} = 1$.

\begin{lemma}[Olver's asymptotics \cite{Watson}]
If $\nu \gg 1$, then $j_\nu = \nu (1 + c \nu^{-2/3} + O(\nu^{-4/3}))$,
with $c = -a_1 2^{-1/3} \approx 1.855757$, where $a_1$ is the first
zero of the Airy function.
\end{lemma}
\begin{corollary}
If $\phi_1 \ll 1$, then
\begin{equation}
\mu = \frac{\alpha_1^2 (1 + \ve_0)^2}{R_2^2} ,
\end{equation}  
with $\alpha_1 = \pi/\phi_1$ and $\ve_0 \propto \alpha_1^{-2/3} \ll 1$.
\end{corollary}

\begin{theorem}
If 
\begin{equation}
\label{eq:Cip_cond}
\phi_1 \ll 1  \quad \mbox{and} \quad  R_1 \ll R_2,
\end{equation}
then there exists an eigenvalue $\lambda$ of the Dirichlet Laplacian
in $\Omega$ which is close to the first eigenvalue $\mu$ of the
Dirichlet Laplacian in $\Omega_3$ from (\ref{eq:Cip_mu}).  In other
words, there exists $\lambda$ such that
\begin{equation}
\label{eq:Cip_lambda}
\lambda = \frac{\alpha_1^2 (1 + \ve')^2}{R_2^2} ,
\end{equation}
with some $\ve' \ll 1$.
\end{theorem} \\
{\bf Proof.}  When $\phi_1$ is small but $R_2$ is large, the
eigenfunction $v$ of the Dirichlet Laplacian in $\Omega_3$ is very
small for $r < R_1$, i.e., in $\Omega_3 \cap \Omega_1$.  This function
is a natural candidate to prove, using Lemma \ref{lem:Av_eps}, the
existence of an eigenvalue $\lambda$ for which the associated
eigenfunction would be localized in $\Omega_2$.

In order to apply Lemma \ref{lem:Av_eps}, we introduce a cut-off
function $\bar \eta(r,\phi) = \eta(r)$ in $\Omega$, with $\eta(r)$
being an analytic function on $\R_+$ such that $\eta(r) = 0$ for $r <
R_1$ and $\eta(r) = 1$ for $r > R_1+\delta$, for a small fixed $\delta
> 0$.  We need to prove that for some small $\ve>0$
\begin{equation}
\label{eq:Cip_auxil22}
\| \Delta( \bar\eta \, v) + \mu \, \bar\eta \, v \|_{L_2(\Omega)} < \ve \| \bar \eta \, v \|_{L_2(\Omega)}.
\end{equation}

First we estimate the left-hand side:
\begin{equation}
\| \Delta( \bar\eta \, v) + \mu \, \bar\eta \, v \|_{L_2(\Omega)} = \| v\, \Delta \bar\eta  + 2 (\nabla \bar\eta \cdot \nabla v) \|_{L_2(Q_\delta)} ,
\end{equation}
where we used that $\{\mu, v\}$ is an eigenpair in $\Omega_3$ while
$\nabla \bar \eta$ is zero everywhere except $Q_\delta = \{ R_1 < r <
R_1 + \delta,~ 0 < \phi < \phi_1\}$.  Since $\bar\eta$ is analytic,
its derivatives are bounded so that
\begin{equation*}
\begin{split}
\| v\, \Delta \bar\eta  + 2 (\nabla \bar\eta \cdot \nabla v) \|^2_{L_2(Q_\delta)} 
& \leq  2\| v\, \Delta \bar\eta \|^2_{L_2(Q_\delta)} + 2 \|2 (\nabla \bar\eta \cdot \nabla v) \|^2_{L_2(Q_\delta)}  \\
& \leq  C_1 \int\limits_{R_1}^{R_1+\delta} dr \, r \, [J_{\alpha_1}(\sqrt{\mu}\,r)]^2 
      + C_2 \int\limits_{R_1}^{R_1+\delta} dr \, r \, [J'_{\alpha_1}(\sqrt{\mu}\,r)]^2 , \\
\end{split}
\end{equation*}
with some positive constants $C_1$ and $C_2$.  Since $\sqrt{\mu}\, r =
\alpha_1(1+\ve_0)r/R_2 \ll \alpha_1$ for $r \in (R_1,R_1+\delta)$ when $R_2
\gg R_1+\delta$, one can apply the asymptotic formula for Bessel functions,
$J_\nu(z) \simeq (z/2)^\nu/\Gamma(\nu+1)$, to estimate the first
integral:
\begin{eqnarray}  \nonumber
C_1 \int\limits_{R_1}^{R_1+\delta} dr \, r \, [J_{\alpha_1}(\sqrt{\mu}\,r)]^2 
&\leq& C_1 \, \delta (R_1 + \delta/2) ~ \frac{(\frac12 \sqrt{\mu} (R_1+\delta))^{2\alpha_1}}{\Gamma^2(\alpha_1+1)}  \\
&\leq& C'_1 \alpha_1 \left(\frac{e(1+\ve_1) (R_1+\delta)}{2R_2}\right)^{2\alpha_1} , 
\end{eqnarray}
which vanishes exponentially fast when $\alpha_1 = \pi/\phi_1$ grows
(here we used the Stirling's formula for the Gamma function).  A
similar estimate holds for the second integral.  We conclude that when
$R_2 \gg R_1$ and $\alpha_1$ is large enough, the left-hand side of
(\ref{eq:Cip_auxil22}) can be made arbitrarily small.

On the other hand, the norm in the right-hand side of
(\ref{eq:Cip_auxil22}) is estimated as
\begin{eqnarray} \nonumber
\| \bar \eta \, v \|^2_{L_2(\Omega)} &=& \| \bar \eta \, v \|^2_{L_2(\Omega_2)} 
= 1 - \| \sqrt{1 - \bar \eta^2} \, v \|^2_{L_2(\Omega_2)} = 1 - \| \sqrt{1 - \bar \eta^2} \, v \|^2_{L_2(Q_\delta)} \\
&\geq& 1 -  \delta (R_1+\delta/2) \, \phi_1  \, \max\limits_{r\in(R_1,R_1+\delta)} \{J_{\alpha_1}^2(\sqrt{\mu}\, r)\} . 
\end{eqnarray}
As a consequence, this norm can be made arbitrarily close to $1$ that
completes the proof of the inequality (\ref{eq:Cip_auxil22}).
According to Lemma \ref{lem:Av_eps}, there exists an eigenvalue
$\lambda$ which is close to $\mu$ that can be written in the form
(\ref{eq:Cip_lambda}), with a new small parameter $\ve'$.  \qed

\subsection{Estimate of the norms}  \label{sec:Cip_estimates}

To prove the localization in the subdomain $\Omega_2$, we estimate the
ratio of $L_2$-norms in two arbitrary ``radial cross-sections'' of
subdomains $\Omega_1$ and $\Omega_2$.  More precisely, we consider the
squared $L_2$-norm of $u_1$ on a circle of radius $r$ inside
$\Omega_1$
\begin{eqnarray} \nonumber
I_1(r) &=& \int\limits_0^{2\pi} d\phi ~ |u_1(r,\phi)|^2 
= \frac{1}{2 \pi} \frac{J_0^2(\sqrt{\lambda}\, r)}{J_0^2(\sqrt{\lambda}\, R_1)} \bigl(u_{\Gamma}, 1\bigr)^2_{L_2(\Gamma)} \\
\label{eq:u1_normG}
&+& \frac{1}{\pi} \sum \limits_{n=1}^{\infty} \frac{J_n^2(\sqrt{\lambda}\, r)}{J_n^2(\sqrt{\lambda}\, R_1)} 
\biggl( \bigl(u_{|\Gamma}, \cos n \phi \bigr)^2_{L_2(\Gamma)} + \bigl( u_{|\Gamma}, \sin n \phi \bigr)^2_{L_2(\Gamma)}\biggr) ,
\end{eqnarray}
and the squared $L_2$-norm of $u_2$ on an arc $(0,\phi_1)$ of radius
$r$ in $\Omega_2$
\begin{equation}
\label{eq:u2_normG}
I_2(r) = \int\limits_0^{\phi_1} d\phi ~ |u_2(r,\phi)|^2 
= \frac{2}{\phi_1}  \sum\limits_{n=1}^{\infty} \frac{\psi_n^2(\sqrt{\lambda}\, r)}{\psi_n^2(\sqrt{\lambda}\, R_1)}
\bigl(u_{|\Gamma}, \sin \alpha_n \phi \bigr)^2_{L_2(\Gamma)} .
\end{equation}
We aim at showing the ratio $I_2(r_2)/I_1(r_1)$ can be made arbitrarily
large as $\phi_1\to 0$.

\begin{theorem}
For any $0 < r_1 < R_1 < r_2 < R_2$, the ratio $I_2(r_2)/I_1(r_1)$ is
bounded from below as
\begin{equation}
\label{eq:Cip_ratio}
\frac{I_2(r_2)}{I_1(r_1)} \geq \frac{\psi_1^2(\sqrt{\lambda}\, r_2)} {\psi_1^2(\sqrt{\lambda}\, R_1)} ~
\frac{J_0^2(\sqrt{\lambda}\, R_1)}{1 + \Psi(\sqrt{\lambda}\, R_1)} ,
\end{equation}
where
\begin{equation}
\label{eq:Psi_disk}
\Psi(r) = - \left(\frac{\psi'_1(r)}{\psi_1(r)} - \frac{J'_0(r)}{J_0(r)} \right) \left(\frac{\psi'_2(r)}{\psi_2(r)} - \frac{J'_0(r)}{J_0(r)}\right)^{-1} .
\end{equation}
\end{theorem} \\
{\bf Proof.} 
First, we rewrite the dispersion relation (\ref{eq:dispersion_disk})
as
\begin{eqnarray} \nonumber
&& \frac{2}{\phi_1} \frac{\psi'_1(\sqrt{\lambda}\, R_1)}{\psi_1(\sqrt{\lambda}\, R_1)} 
\bigl(u_{|\Gamma}, \sin \alpha_1 \phi \bigr)^2_{L_2(\Gamma)} -
\frac{1}{2 \pi} \frac{J'_0(\sqrt{\lambda}\, R_1)}{J_0(\sqrt{\lambda}\, R_1)} \bigl(u_{|\Gamma}, 1\bigr)^2_{L_2(\Gamma)} \\
\nonumber
&& = \frac{1}{\pi} \sum\limits_{n=1}^{\infty}
\frac{J'_n(\sqrt{\lambda}\, R_1)}{J_n(\sqrt{\lambda}\, R_1)} \biggl(\bigl( u_{|\Gamma}, \cos n \phi \bigr)^2_{L_2(\Gamma)} +
\bigl(u_{|\Gamma}, \sin n \phi\bigr)^2_{L_2(\Gamma)} \biggr)  \\
&& - \frac{2}{\phi_1} \sum\limits_{n=2}^{\infty} \frac{\psi'_n(\sqrt{\lambda}\, R_1)}{\psi_n(\sqrt{\lambda}\, R_1)} 
\bigl(u_{|\Gamma}, \sin \alpha_n \phi \bigr)^2_{L_2(\Gamma)}  .
\end{eqnarray}
The inequality (\ref{eq:J_ineq3}) ensures that the first term in the
right-hand side is positive, from which
\begin{eqnarray} \nonumber
&& \frac{2}{\phi_1} \frac{\psi'_1(\sqrt{\lambda}\, R_1)}{\psi_1(\sqrt{\lambda}\, R_1)} 
\bigl(u_{|\Gamma}, \sin \alpha_1 \phi \bigr)^2_{L_2(\Gamma)} -
\frac{1}{2 \pi} \frac{J'_0(\sqrt{\lambda}\, R_1)}{J_0(\sqrt{\lambda}\, R_1)} \bigl(u_{|\Gamma}, 1\bigr)^2_{L_2(\Gamma)} \\
&& \geq - \frac{2}{\phi_1} \sum\limits_{n=2}^{\infty} \frac{\psi'_n(\sqrt{\lambda}\, R_1)}{\psi_n(\sqrt{\lambda}\, R_1)} 
\bigl(u_{|\Gamma}, \sin \alpha_n \phi \bigr)^2_{L_2(\Gamma)}  .
\end{eqnarray}
Expressions (\ref{eq:u1_normG}, \ref{eq:u2_normG}) at $r = R_1$ imply
\begin{equation}
\frac{1}{2 \pi} \bigl(u_{|\Gamma}, 1 \bigr)^2_{L_2(\Gamma)} \leq I_1(R_1) = I_2(R_1) =
\frac{2}{\phi_1} \sum \limits_{n=1}^{\infty} \bigl(u_{|\Gamma}, \sin \alpha_n \phi \bigr)^2_{L_2(\Gamma)}  ,
\end{equation}
so that
\begin{eqnarray*} 
&& - \frac{2}{\phi_1} \sum\limits_{n=2}^{\infty} \frac{\psi'_n(\sqrt{\lambda}\, R_1)}{\psi_n(\sqrt{\lambda}\, R_1)} 
\bigl(u_{|\Gamma}, \sin \alpha_n \phi \bigr)^2_{L_2(\Gamma)} \\
&& \leq \frac{2}{\phi_1} \frac{\psi'_1(\sqrt{\lambda}\, R_1)}{\psi_1(\sqrt{\lambda}\, R_1)} 
\bigl(u_{|\Gamma}, \sin \alpha_1 \phi \bigr)^2_{L_2(\Gamma)} -
\frac{J'_0(\sqrt{\lambda}\, R_1)}{J_0(\sqrt{\lambda}\, R_1)} 
\frac{2}{\phi_1} \sum \limits_{n=1}^{\infty} \bigl(u_{|\Gamma}, \sin \alpha_n \phi \bigr)^2_{L_2(\Gamma)} ,
\end{eqnarray*}
where we used the inequality
\begin{equation}
\frac{J'_0(\sqrt{\lambda}\, R_1)}{J_0(\sqrt{\lambda}\, R_1)} = - \sqrt{\lambda} \frac{J_1(\sqrt{\lambda}\, R_1)}{J_0(\sqrt{\lambda}\, R_1)} < 0  
\end{equation}
that follows from (\ref{eq:J_ineq0}).

As a result, we get
\begin{eqnarray}  \nonumber
&& - \sum\limits_{n=2}^{\infty} \biggl(\frac{\psi'_n(\sqrt{\lambda}\, R_1)}{\psi_n(\sqrt{\lambda}\, R_1)}  - 
\frac{J'_0(\sqrt{\lambda}\, R_1)}{J_0(\sqrt{\lambda}\, R_1)} \biggr) \bigl(u_{|\Gamma}, \sin \alpha_n \phi \bigr)^2_{L_2(\Gamma)} \\
\label{eq:auxil2}
&& \leq \biggl(\frac{\psi'_1(\sqrt{\lambda}\, R_1)}{\psi_1(\sqrt{\lambda}\, R_1)} - \frac{J'_0(\sqrt{\lambda}\, R_1)}{J_0(\sqrt{\lambda}\, R_1)} \biggr)
\bigl(u_{|\Gamma}, \sin \alpha_1 \phi \bigr)^2_{L_2(\Gamma)} .
\end{eqnarray}
The inequality (\ref{eq:psi_ineq}) with $n_1 = 2$ and $n_2 = n \geq 2$
yields
\begin{equation}
\label{eq:auxil5}
- \frac{\psi'_n(\sqrt{\lambda}\, R_1)}{\psi_n(\sqrt{\lambda}\, R_1)} \geq - \frac{\psi'_2(\sqrt{\lambda}\, R_1)}{\psi_2(\sqrt{\lambda}\, R_1)} ,
\end{equation}
from which one deduces
\begin{eqnarray}  \nonumber
&& - \biggl(\frac{\psi'_2(\sqrt{\lambda}\, R_1)}{\psi_2(\sqrt{\lambda}\, R_1)} - \frac{J'_0(\sqrt{\lambda}\, R_1)}{J_0(\sqrt{\lambda}\, R_1)}\biggr) 
\sum \limits_{n=2}^{\infty}  \bigl(u_{|\Gamma}, \sin \alpha_n \phi \bigr)^2_{L_2(\Gamma)}  \\
\label{eq:auxil3}
&& \leq \biggl(\frac{\psi'_1(\sqrt{\lambda}\, R_1)}{\psi_1(\sqrt{\lambda}\, R_1)} - \frac{J'_0(\sqrt{\lambda}\, R_1)}{J_0(\sqrt{\lambda}\, R_1)}\biggr)
 \bigl(u_{|\Gamma}, \sin \alpha_1 \phi \bigr)^2_{L_2(\Gamma)} .
\end{eqnarray}
Using the inequality (\ref{eq:psi2_J0}), we rewrite it as
\begin{equation}
\label{eq:auxil3a}
\sum \limits_{n=2}^{\infty}  \bigl(u_{|\Gamma}, \sin \alpha_n \phi \bigr)^2_{L_2(\Gamma)} \leq 
\Psi(\sqrt{\lambda}\, R_1) \, \bigl(u_{|\Gamma}, \sin \alpha_1 \phi \bigr)^2_{L_2(\Gamma)} ,
\end{equation}
where $\Psi(r)$ is defined in (\ref{eq:Psi_disk}).

With these inequalities, we estimate the squared $L_2$-norm from
(\ref{eq:u1_normG}):
\begin{eqnarray*}
I_1(r_1) &\leq& \frac{1}{J_0^2(\sqrt{\lambda}\, R_1)} \frac{1}{2 \pi} \bigl(u_{|\Gamma}, 1\bigr)^2_{L_2(\Gamma)} 
+ \frac{1}{\pi} \sum\limits_{n=1}^{\infty}  \biggl( \bigl( u_{|\Gamma}, \cos n \phi \bigr)^2_{L_2(\Gamma)} +
\bigl( u_{|\Gamma}, \sin n \phi \bigr)^2_{L_2(\Gamma)} \biggr) \\
&\leq&  \frac{1}{J_0^2(\sqrt{\lambda}\, R_1)} I_1(R_1) \\
&=& \frac{1}{J_0^2(\sqrt{\lambda}\, R_1)} I_2(R_1) \\
&=& \frac{1}{J_0^2(\sqrt{\lambda}\, R_1)} \frac{2}{\phi_1} 
\biggl[\bigl(u_{|\Gamma}, \sin \alpha_1 \phi \bigr)^2_{L_2(\Gamma)} + \sum \limits_{n=2}^{\infty}  \bigl(u_{|\Gamma}, \sin \alpha_n \phi \bigr)^2_{L_2(\Gamma)} \biggr] \\
&\leq & \frac{1}{J_0^2(\sqrt{\lambda}\, R_1)}  \frac{2}{\phi_1} \biggl[ \bigl(u_{|\Gamma}, \sin \alpha_1 \phi\bigr)^2_{L_2(\Gamma)} + 
\Psi(\sqrt{\lambda}\, R_1)\, \bigl(u_{|\Gamma}, \sin \alpha_1 \phi \bigr)^2_{L_2(\Gamma)} \biggr] \\
&=& \frac{1 + \Psi(\sqrt{\lambda}\, R_1)}{J_0^2(\sqrt{\lambda}\, R_1)} 
\frac{2}{\phi_1} \bigl(u_{|\Gamma}, \sin \alpha_1 \phi \bigr)^2_{L_2(\Gamma)} ,
\end{eqnarray*}
where we used the inequality (\ref{eq:psi2_J0}).

On the other hand, we obtain
\begin{equation}
\label{eq:u2_normG_ineq}
I_2(r_2) \geq \frac{2}{\phi_1} \frac{\psi_1^2(\sqrt{\lambda}\, r_2)}{\psi_1^2(\sqrt{\lambda}\, R_1)} \bigl(u_{|\Gamma}, \sin \alpha_1 \phi\bigr)^2_{L_2(\Gamma)} .
\end{equation}
Combining these inequalities, we conclude for
any $0\leq r_1 \leq R_1 \leq r_2 \leq R_2$ that
\begin{equation}
\frac{I_2(r_2)}{I_1(r_1)} \geq \frac{\psi_1^2(\sqrt{\lambda}\, r_2)} {\psi_1^2(\sqrt{\lambda}\, R_1)} ~
\frac{J_0^2(\sqrt{\lambda}\, R_1)}{1 + \Psi(\sqrt{\lambda}\, R_1)} 
\end{equation}
that completes the proof of the theorem. \qed

\begin{lemma} \label{lem:psi1_large}
Under conditions (\ref{eq:Cip_cond}), if $|\lambda - \mu| < \ve$ for
some $\ve > 0$, then $|\psi_1(\sqrt{\lambda}\, R_1)| < C \ve$.
\end{lemma} \\
{\bf Proof.}  The continuity of Bessel functions implies
\begin{equation}
\lim\limits_{\lambda\to \mu} \psi_1(\sqrt{\lambda}\, R_1) = \psi_1(\sqrt{\mu}\, R_1) = J_{\alpha_1}(\sqrt{\mu}\, R_1) \, Y_{\alpha_1}(\sqrt{\mu}\, R_2),
\end{equation}
where the second term in the definition (\ref{eq:psin}) was dropped
because $\mu$ is the Dirichlet eigenvalue for the sector $\Omega_3$
and thus $J_{\alpha_1}(\sqrt{\mu}\, R_2) = 0$.

When $R_1 \ll R_2$, one can apply the asymptotic formulas for the
Bessel functions
\begin{eqnarray}
J_{\alpha_1}(\sqrt{\mu}\, R_1) &\simeq& \frac{1}{\Gamma(\alpha_1+1)} ~ \left(\frac{\alpha_1(1+\ve_1) R_1}{2R_2}\right)^{\alpha_1}  , \\
Y_{\alpha_1}(\sqrt{\mu}\, R_2) &\simeq& - \frac{\Gamma(\alpha_1+1)}{\pi} ~ \left(\frac{2}{\alpha_1(1+\ve_1)}\right)^{\alpha_1}, 
\end{eqnarray}
from which
\begin{equation}
|\psi_1(\sqrt{\mu}\, R_1)| \simeq \frac{1}{\pi} \left(\frac{R_1}{R_2}\right)^{\alpha_1}.
\end{equation}
When $R_1 \ll R_2$ and $\alpha_1$ is large, the right-hand side can be
made arbitrarily small that completes the proof.  \qed

\begin{corollary}
Under conditions (\ref{eq:Cip_cond}), there exists an eigenfunction
$u$ of the Dirichlet Laplacian in $\Omega$ which is localized in
$\Omega_2$.  In particular, the ratio $I_2(r_2)/I_1(r_1)$ of squared
$L_2$ norms from (\ref{eq:Cip_ratio}) can be made arbitrarily large.
\end{corollary} \\
{\bf Proof.}  Let us examine the right-hand side of
(\ref{eq:Cip_ratio}).  The function $\psi_1(\sqrt{\lambda}\, r_2)$
does not vanish on $R_1 < r_2 < R_2$ except at $r_2 = R_2$.  Since
$J_0(\sqrt{\lambda}\, R_1)$ is a constant, it remains to show that
$\psi_1^2(\sqrt{\lambda}\, R_1) (1+ \Psi(\sqrt{\lambda}\, R_1))$ can
be made arbitrarily small.  On one hand, the absolute value of
\begin{eqnarray} \nonumber
\psi_1(\sqrt{\lambda}\, R_1) \bigl(1+ \Psi(\sqrt{\lambda}\, R_1) \bigr)
&=& \psi_1(\sqrt{\lambda}\, R_1) - \left(\psi'_1(\sqrt{\lambda}\, R_1) 
- \psi_1(\sqrt{\lambda}\, R_1) \frac{J'_0(\sqrt{\lambda}\, R_1)}{J_0(\sqrt{\lambda}\, R_1)} \right)  \\
&\times& \left(\frac{\psi'_2(\sqrt{\lambda}\, R_1)}{\psi_2(\sqrt{\lambda}\, R_1)} - \frac{J'_0(\sqrt{\lambda}\, R_1)}{J_0(\sqrt{\lambda}\, R_1)}\right)^{-1} 
\end{eqnarray}
can be bounded from above by a constant.  On the other hand, the
remaining factor $\psi_1(\sqrt{\lambda}\, R_1)$ can be made
arbitrarily small by Lemma \ref{lem:psi1_large} that completes the
proof.  \qed

\appendix
\section{Numerical illustrations}
\label{sec:numerics}

We illustrate the localization of eigenfunctions by considering a
rectangle $[-a_2,a_1]\times [0,1]$ with a vertical slit $[h,1]$:
$\Omega = ([-a_1,a_2] \times [0,1]) \backslash (\{0\}\times [0,h])$,
i.e., two rectangles $[-a_1,0]\times [0,1]$ and $[0,a_2]\times [0,1]$
connected through an opening $\Gamma = [0,h]$ at $x = 0$
(Fig. \ref{fig:domain}a).  Setting $a_1 = 1$ and $a_2 = 0.8$, we
compute several eigenfunctions of the Dirichlet Laplacian by a finite
element method in Matlab PDEtools for several values of $h$.

\begin{figure}
\begin{center}
\includegraphics[width=90mm]{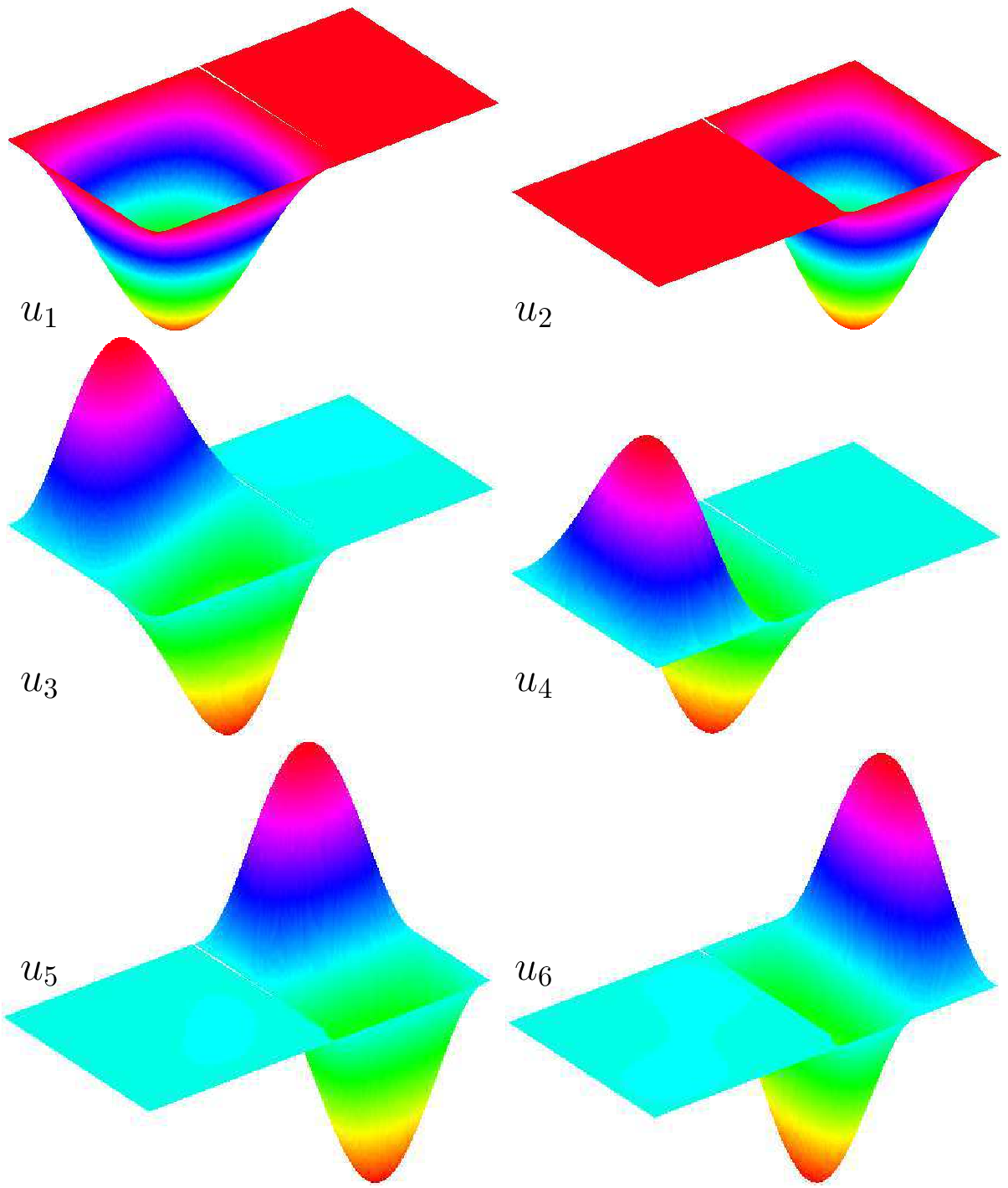}  % fig_slit1_.eps}
\end{center}
\caption{
The first six Dirichlet eigenfunctions in a rectangle with a vertical
slit, $\Omega = ([-a_1,a_2] \times [0,1]) \backslash (\{0\}\times
[0,h])$, with $a_1 = 1$, $a_2 = 0.8$ and $h = 0.1$.  Eigenfunctions
$u_1$, $u_3$, $u_4$ are localized in the larger domain $\Omega_1$
while $u_2$, $u_5$ and $u_6$ are localized in $\Omega_2$. }
\label{fig:eigen_1BD_h01}
\end{figure}

\begin{figure}
\begin{center}
\includegraphics[width=90mm]{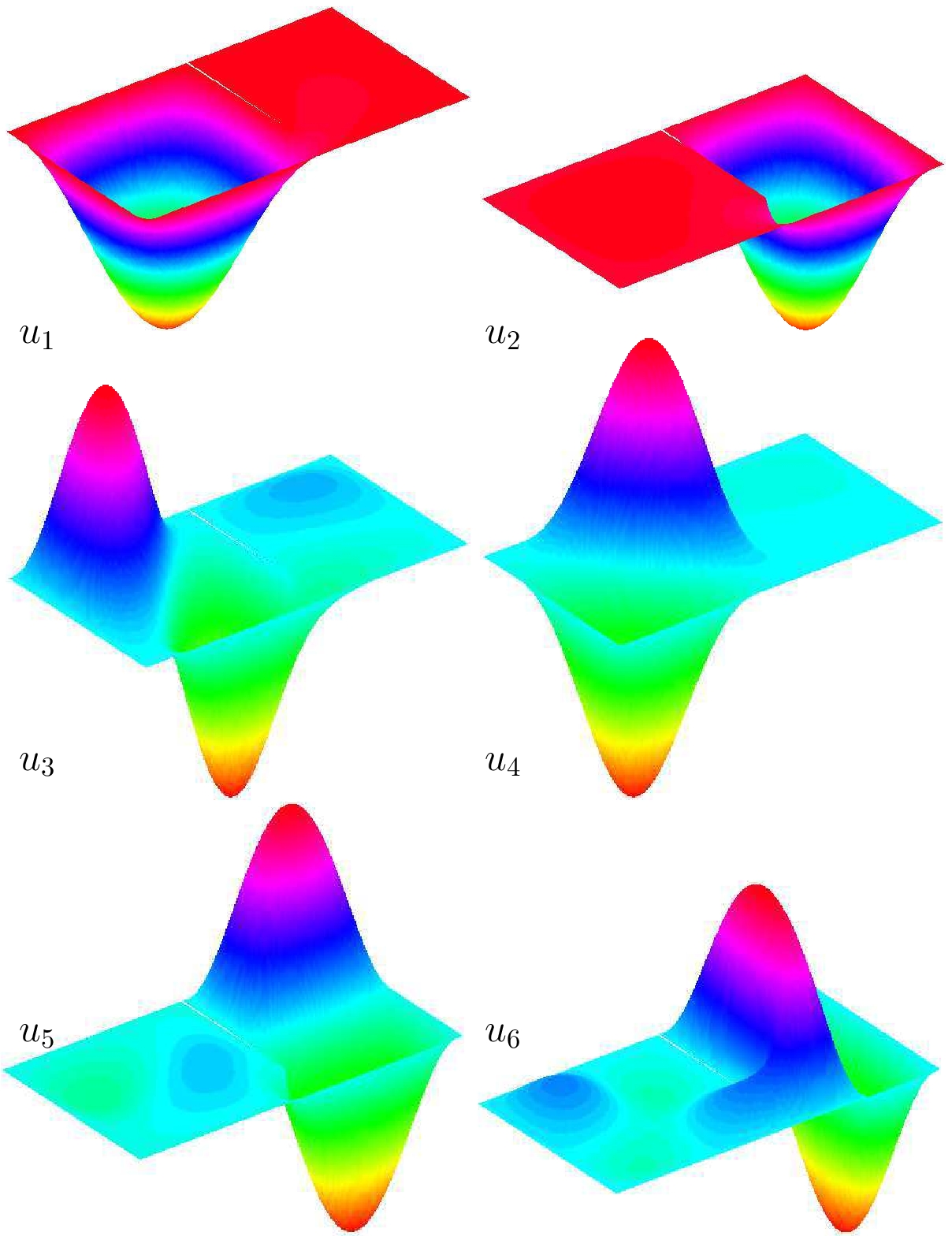} % fig_slit2_.eps}
\end{center}
\caption{
The first six Dirichlet eigenfunctions in a rectangle with a slit,
$\Omega = ([-a_1,a_2] \times [0,1]) \backslash (\{0\}\times [0,h])$,
with $a_1 = 1$, $a_2 = 0.8$ and $h = 0.25$.  Eigenfunctions $u_1$,
$u_3$, $u_4$ are localized in the larger domain $\Omega_1$ while
$u_2$, $u_5$ and $u_6$ are localized in $\Omega_2$. }
\label{fig:eigen_1BD_h025}
\end{figure}

\begin{figure}
\begin{center}
\includegraphics[width=90mm]{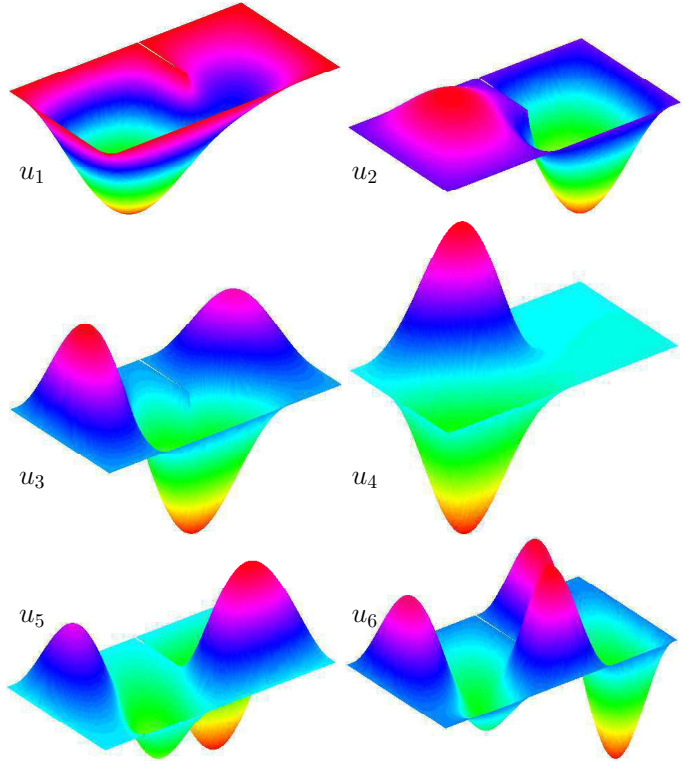}  % fig_slit3_.eps}
\end{center}
\caption{
The first six Dirichlet eigenfunctions in a rectangle with a slit,
$\Omega = ([-a_1,a_2] \times [0,1]) \backslash (\{0\}\times [0,h])$,
with $a_1 = 1$, $a_2 = 0.8$ and $h = 0.5$.  Only eigenfunctions $u_1$
and $u_4$ remain localized. }
\label{fig:eigen_1BD_h05}
\end{figure}

Figure \ref{fig:eigen_1BD_h01} shows the first six Dirichlet
eigenfunctions for $h = 0.1$.  Even though the diameter of the opening
$\Gamma$ is not so small, one observes a very strong localization:
eigenfunctions $u_1$, $u_3$, $u_4$ are localized in the larger domain
$\Omega_1$, whereas $u_2$, $u_5$ and $u_6$ are localized in
$\Omega_2$.  The corresponding eigenvalues are provided in Table
\ref{tab:eigenvalues}.  Even if the opening $\Gamma$ is increased to
the quarter of the rectangle width ($h = 0.25$), the localization of
first eigenfunctions is still present (Fig. \ref{fig:eigen_1BD_h025}).
However, one can see that the eigenfunctions localized in one
subdomain start to penetrate into the other subdomain.  This
penetration is enhanced for higher-order eigenfunctions.  Setting $h =
0.5$ destroys the localization of eigenfunctions, except for $u_1$ and
$u_4$ (Fig. \ref{fig:eigen_1BD_h05}).  Looking at these figures in the
backward order, one can see the progressive emergence of the
localization as the opening $\Gamma$ shrinks.  It is remarkable how
strong the localization can be even for not too narrow openings.

\begin{figure}
\begin{center}
\includegraphics[width=60mm]{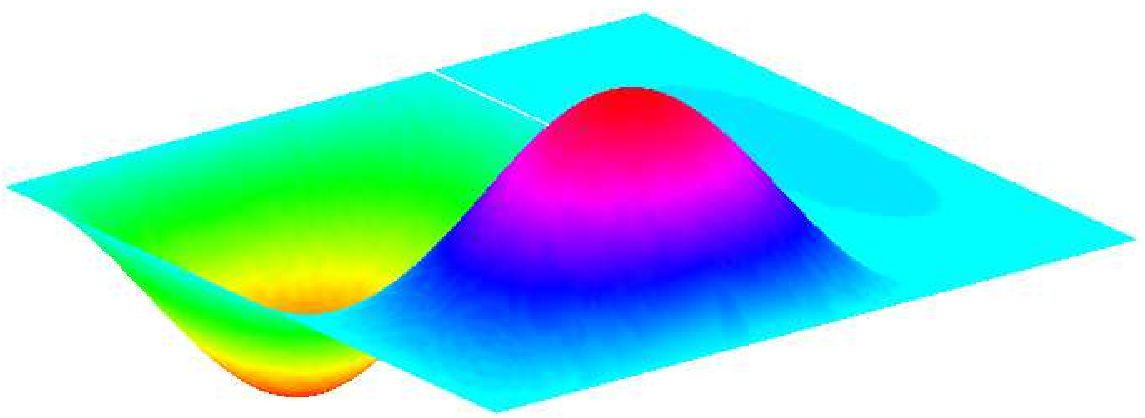} % figures/rectangle_a1_1_a2_05_h01_eigen2D.eps}
\end{center}
\caption{
The second Dirichlet eigenfunction in a rectangle with a vertical
slit, $\Omega = ([-a_1,a_2] \times [0,1]) \backslash (\{0\}\times
[0,h])$, with $a_1 = 1$, $a_2 = 0.5$ and $h = 0.1$.  This
eigenfunction is still localized in $\Omega_1$ although the condition
(\ref{eq:1BD_auxil6}) is not satisfied: $\sqrt{a_2^2(\nu_2 -
\nu_1)/\pi^2 + a_2^2/a_1^2} = 1$, given that $\nu_1 = \pi^2$ and
$\nu_2 = 4\pi^2$.}
\label{fig:eigen_1BD2}
\end{figure}

Finally, Fig. \ref{fig:eigen_1BD2} shows the second eigenfunction for
the geometric setting with $a_1 = 1$ and $a_2 = 0.5$.  For this choice
of $a_2$, the condition (\ref{eq:1BD_auxil6}), which was used to prove
the localization of the second eigenfunction in
Sec. \ref{sec:1BD_other}, is not satisfied.  Nevertheless, the second
eigenfunction turns out to be localized in the larger domain.  This
example suggests that the condition (\ref{eq:1BD_auxil6}) may
potentially be relaxed.

\begin{table}
\begin{center}
\begin{tabular}{| c | r | c | c | c | l |}  \hline
$n$ & $h = 0$                          & $h = 0.1$ & $h = 0.25$ & $h = 0.5$ & $h = 1$ \\  \hline
1   & {\small $(1+1)\pi^2 \approx$} $19.74$       &  $19.79$  &  $19.70$   &  $18.39$  & $12.92$ {\small $\approx \pi^2(1 + 1/1.8^2)$} \\
2   & {\small $(1+1/0.8^2)\pi^2 \approx$} $25.29$ &  $25.39$  &  $25.22$   &  $23.58$  & $22.05$ {\small $\approx \pi^2(1 + 4/1.8^2)$} \\
3   & {\small $(1+4)\pi^2\approx$} $49.35$        &  $49.42$  &  $48.77$   &  $41.40$  & $37.29$ {\small $\approx \pi^2(1 + 9/1.8^2)$} \\
4   & {\small $(1+4)\pi^2\approx$} $49.35$        &  $49.59$  &  $49.49$   &  $49.33$  & $42.52$ {\small $\approx \pi^2(4 + 1/1.8^2)$} \\
5   & {\small $(4+1/0.8^2)\pi^2\approx$} $54.30$  &  $55.04$  &  $54.52$   &  $53.32$  & $51.66$ {\small $\approx \pi^2(4 + 4/1.8^2)$} \\
6   & {\small $(1+4/0.8^2)\pi^2\approx$} $71.55$  &  $72.01$  &  $70.99$   &  $62.88$  & $58.61$ {\small $\approx \pi^2(1 + 16/1.8^2)$} \\  \hline
\end{tabular}
\end{center}
\caption{
The first six Dirichlet eigenvalues in a rectangle with a vertical
slit, $\Omega = ([-a_1,a_2] \times [0,1]) \backslash (\{0\}\times
[0,h])$, with $a_1 = 1$, $a_2 = 0.8$, and five values of $h$,
including the limit cases: $h = 0$ (no opening, two disjoint
subdomains) and $h = 1$ (no barrier).  }
\label{tab:eigenvalues}
\end{table}

% mpoly = load('c:\dg\matlab\polygons\eigen_rectangle_a1_1_a2_0.8_h_0.1.txt', '-ASCII');

% mpoly = load('c:\dg\matlab\polygons\eigen_rectangle_a1_1_a2_0.8_h_0.25.txt', '-ASCII');

% mpoly = load('c:\dg\matlab\polygons\eigen_rectangle_a1_1_a2_0.8_h_0.5.txt', '-ASCII');

\section{Proof of the classical lemma \ref{lem:Av_eps}}
\label{sec:proof_lemma}

Here we provide an elementary proof of the classical lemma
\ref{lem:Av_eps}.

{\bf Proof.}  Let $\lambda_k$ and $\psi_k$ denote eigenvalues and
$L_2$-normalized eigenfunctions of $A$ forming a complete basis in
$L_2$.  Let us decompose $v$ on this basis: $v = \sum_k c_k \psi_k$.
One has then
\begin{equation}
\| A v - \mu v \|^2_{L_2} = \biggl\| \sum\limits_k (\lambda_k - \mu) c_k \psi_k \biggr\|^2_{L_2} = \sum\limits_k (\lambda_k - \mu)^2 c_k^2 
\geq \min\limits_k \{ (\lambda_k - \mu)^2 \}  \sum\limits_k c_k^2 ,
\end{equation}
i.e.,
\begin{equation}
\| A v - \mu v \|_{L_2} \geq \min\limits_k \{ |\lambda_k - \mu| \} ~ \| v\|_{L_2} .
\end{equation}
On the other hand, (\ref{eq:Av_eps}) implies that 
\begin{equation}
\min\limits_k \{ |\lambda_k - \mu| \} < \ve
\end{equation}
that completes the proof.  \qed

\section{Some inequalities involving Bessel functions}  \label{sec:Cip_inequalities}

In this Appendix, we prove several inequalities involving Bessel
functions.  The technique of proofs is standard and can be found in
classical textbooks \cite{Watson,Bowman,Abramowitz}. \\
\begin{lemma} \label{lem:Jineq0}
If $\sqrt{\lambda}\, R_1 \leq j'_1$, then for any $\nu \geq 0$,
\begin{equation}
\label{eq:J_ineq0}
J_{\nu}(\sqrt{\lambda}\, r) > 0 \qquad \forall ~ 0 < r \leq R_1,
\end{equation}
where $j'_1 \approx 1.8412$ is the first zero of $J'_1(z)$.
\end{lemma} \\
{\bf Proof.}  The known inequalities on the first zeros $j_\nu$ of
Bessel functions $J_\nu(z)$,
\begin{equation}
j'_1 < j_0 \leq j_\nu  \qquad \forall ~ \nu \geq 0,
\end{equation}
imply that $J_\nu(\sqrt{\lambda}\, r)$ does not change sign in the
interval $(0,R_1)$.  In turn, the asymptotic behavior $J_\nu(x) \simeq
(x/2)^\nu/\Gamma(\nu+1)$ as $x \to 0$ ensures the positive sign. \qed

\begin{lemma} \label{lem:Jineq1}
If $\sqrt{\lambda}\, R_1 \leq j'_1$, then for any $0 \leq \nu_1 <
\nu_2$, one has
\begin{equation}
\label{eq:J_ineq1}
\frac{J'_{\nu_1}(\sqrt{\lambda}\, r)}{J_{\nu_1}(\sqrt{\lambda}\, r)} 
\leq \frac{J'_{\nu_2}(\sqrt{\lambda}\, r)}{J_{\nu_2}(\sqrt{\lambda}\, r)}  \qquad \forall ~ 0 < r \leq R_1.
\end{equation}
\end{lemma} \\
{\bf Proof.}  Although the proof is standard, we provide it for
completeness.

Writing the Bessel equations for $J_{\nu_1}(\sqrt{\lambda}\, r)$ and
$J_{\nu_2}(\sqrt{\lambda}\, r)$,
\begin{eqnarray*}
\frac{1}{r} \frac{d}{d r} \left(r \frac{d J_{\nu_1}(\sqrt{\lambda}\,r)}{d r}\right) + \lambda J_{\nu_1}(\sqrt{\lambda}\,r) 
- \frac{\nu_1^2}{r^2} J_{\nu_1}(\sqrt{\lambda}r) &=& 0 ,\\
\frac{1}{r} \frac{d}{d r} \left(r \frac{d J_{\nu_2}(\sqrt{\lambda}\,r)}{d r}\right) + \lambda J_{\nu_2}(\sqrt{\lambda}\,r) 
- \frac{\nu_2^2}{r^2} J_{\nu_2}(\sqrt{\lambda}r) &=& 0 ,
\end{eqnarray*}
multiplying the first one by $r J_{\nu_2}(\sqrt{\lambda}\, r)$, the
second one by $r J_{\nu_1}(\sqrt{\lambda}\, r)$, and subtracting one
from the other, one gets
\begin{eqnarray}  \nonumber
&& \frac{d}{d r} r \left(J_{\nu_1}(\sqrt{\lambda}\, r) \frac{d J_{\nu_2}(\sqrt{\lambda}\, r)}{d r} 
 - J_{\nu_2}(\sqrt{\lambda}\, r) \frac{d J_{\nu_1}(\sqrt{\lambda}\, r)}{d r}\right) \\
\label{eq:J_ineq_aux}
&&  + \frac{\nu_1^2 - \nu_2^2}{r} J_{\nu_1}(\sqrt{\lambda}\, r) J_{\nu_2}(\sqrt{\lambda}\, r) = 0 .
\end{eqnarray}
The integration from $0$ to $r$ yields
\begin{eqnarray}  \nonumber
&& \sqrt{\lambda} r \left(J_{\nu_1}(\sqrt{\lambda}\, r) J'_{\nu_2}(\sqrt{\lambda}\, r) - J_{\nu_2}(\sqrt{\lambda}\, r) J'_{\nu_1}(\sqrt{\lambda}\, r)\right) \\
&& = \int\limits_0^r \frac{\nu_2^2 - \nu_1^2}{r'} J_{\nu_1}(\sqrt{\lambda}\, r') J_{\nu_2}(\sqrt{\lambda}\, r') dr' ,
\end{eqnarray}
and the integral is strictly positive due to (\ref{eq:J_ineq0}).  \qed

\begin{lemma}
If $\sqrt{\lambda}\, R_1 \leq j'_1$, then for any $\nu \geq 1$, one
has
\begin{eqnarray}
\label{eq:J_ineq20}
J'_\nu(\sqrt{\lambda}\, r) &\geq&  0  \hskip 21.5mm  \forall~ 0 < r < R_1 ,  \\
\label{eq:J_ineq2}
J_\nu(\sqrt{\lambda}\, r) &\leq& J_\nu(\sqrt{\lambda}\, R_1)  \qquad \forall ~ 0 < r \leq R_1, \\
\label{eq:J_ineq3}
\frac{J'_{\nu}(\sqrt{\lambda}\, r)}{J_{\nu}(\sqrt{\lambda}\, r)} &\geq& 0 \hskip 21.7mm \forall ~ 0 < r \leq R_1 .
\end{eqnarray}
\end{lemma} \\
{\bf Proof.}  The first inequality (\ref{eq:J_ineq20}) is a direct
consequence of Lemma \ref{lem:Jineq0}, Lemma \ref{lem:Jineq1}, and the
inequality $J'_1(\sqrt{\lambda}\, r) \geq 0$ which is fulfilled for $0
< r < R_1$.  The second inequality (\ref{eq:J_ineq2}) follows from the
first one.  The third inequality (\ref{eq:J_ineq3}) is a consequence
of the first one and Lemma \ref{lem:Jineq0}.  \qed

\vskip 2mm
Now we turn to the function $\psi_n(\sqrt{\lambda}\,r)$ defined by
(\ref{eq:psin}) which satisfies the Bessel equation:
\begin{equation}
\label{eq:psin_bessel}
\frac{1}{r} \frac{d}{d r} \left(r \frac{d \psi_{n}(\sqrt{\lambda}\,r)}{d r}\right) + \lambda \psi_{n}(\sqrt{\lambda}\,r) 
- \frac{\alpha_n^2}{r^2} \psi_{n}(\sqrt{\lambda}\, r) = 0 .
\end{equation}

\begin{lemma}
For any $0 < a < b$, one has
\begin{equation}
\label{eq:Cip_Besselnorm}
\int\limits_a^b dr \, r \, \psi_n^2(\sqrt{\lambda}\, r) = \frac{1}{2\lambda} \left.\left( (r \psi'_n(\sqrt{\lambda}\, r))^2 
+ (\lambda r^2 - \alpha_n^2)\psi_n^2(\sqrt{\lambda}\, r) \right)\right|_{r=a}^{r=b} .
\end{equation}
\end{lemma}\\
{\bf Proof.}  The proof is obtained by multiplying the Bessel equation
(\ref{eq:psin_bessel}) by $r^2\psi'_n$ and integrating from $a$ to $b$.
\qed

\begin{lemma}  \label{lem:psin2}
If $\alpha_n^2/R_2^2 > \lambda$, then
\begin{equation}
\label{eq:psin_psi_general}
- \frac{\psi'_n(\sqrt{\lambda}\, r)}{\psi_n(\sqrt{\lambda}\, r)} \geq 
\frac{r \, b(r) \, b(R_2)}{\lambda^{3/2} + \sqrt{\lambda^3 + r^2 \, b(r) \, b^2(R_2)}} \qquad \forall~ 0 < r < R_2,
\end{equation}
where $b(r) = \alpha_n^2/r^2 - \lambda$.
\end{lemma}  \\
{\bf Proof.}  We introduce a new function $v(z) =
\psi_n(\sqrt{\lambda}\, r)$ by changing the variable $z = \ln
(r/\alpha_n)$, with $r \in (0,R_2)$.  Starting from the Bessel equation
for $\psi_n(\sqrt{\lambda}\, r)$, it is easy to check that the new
function $v$ satisfies the equation
\begin{equation}
v''(z) = (1 - \lambda e^{2z}) \alpha^2_n v(z) ,
\end{equation}
where prime denotes the derivative with respect to $z$.  One has then
\begin{eqnarray}  \nonumber
(v^2)'' &=& 2v v'' + 2(v')^2 \geq 2v v'' = 2\alpha^2_n (1 - \lambda e^{2z}) v^2 \\  \nonumber
&=& 2\alpha^2_n \left(1 - \lambda (r/\alpha_n)^2\right) v^2 \geq 2\alpha^2_n \left(1 - \lambda (R_2/\alpha_n)^2\right) v^2 \\
&=& 2R_2^2 \, b(R_2) \, v^2 ,
\end{eqnarray}
where we used $r < R_2$, and $b(R_2) = \alpha_n^2/R_2^2 - \lambda >
0$.  Integration of this inequality from $z$ to $z_2 =
\ln(R_2/\alpha_n)$ yields
\begin{equation}
-2v' v = - (v^2)' = \int\limits_z^{z_2} dz' (v^2)''  \geq 2 R_2^2 \, b(R_2) \int\limits_z^{z_2} dz' v^2(z'),
\end{equation}
where $(v^2)'_{z=z_2} = 2v'(z_2) v(z_2) = 0$ due to the boundary
condition $\psi_n(\sqrt{\lambda}\, R_2) = 0$.  One gets then
\begin{equation}
- \frac{v'(z)}{v(z)} \geq \frac{R_2^2 \, b(R_2)}{v^2(z)} \int\limits_z^{z_2} dz' v^2(z').
\end{equation}
Since 
\begin{equation}
\frac{d}{dr} \psi_n(\sqrt{\lambda}\, r) = \frac{dv}{dr} = \frac{dz}{dr} \frac{dv}{dz} = \frac{1}{r} v',
\end{equation}
one obtains
\begin{equation}
\label{eq:psin_C}
- \frac{\frac{d}{dr} \psi_n(\sqrt{\lambda}\, r)}{\psi_n(\sqrt{\lambda}\, r)} \geq 
\frac{R_2^2 \, b(R_2)}{r \psi_n^2(\sqrt{\lambda}\, r)} \int\limits_r^{R_2} \frac{dr'}{r'} \psi_n^2(\sqrt{\lambda}\, r') .
\end{equation}
We conclude that
\begin{equation}
\label{eq:psin_ineq}
- \frac{\psi'_n(\sqrt{\lambda}\, r)}{\psi_n(\sqrt{\lambda}\, r)} > 0.
\end{equation}

One can further improve the lower bound.  The integral in the
right-hand side of (\ref{eq:psin_C}) can be estimated as
\begin{eqnarray*}
\int\limits_r^{R_2} \frac{dr'}{r'} \psi_n^2(\sqrt{\lambda}\, r') &=& \int\limits_r^{R_2} \frac{dr'}{r'^2} r' \psi_n^2(\sqrt{\lambda}\, r')
\geq \frac{1}{R_2^2} \int\limits_r^{R_2} dr'\, r' \,\psi_n^2(\sqrt{\lambda}\, r') \\  
&=& \frac{1}{R_2^2} \frac{1}{2\lambda} \left. \left( (r' \psi'_n(\sqrt{\lambda}\, r'))^2 
+ (\lambda r'^2 - \alpha_n^2)\psi_n^2(\sqrt{\lambda}\, r') \right)\right|_{r'=r}^{r'=R_2} \\
&=& \frac{1}{2\lambda R_2^2} \left( (R_2 \psi'_n(\sqrt{\lambda}\, R_2))^2 - (r \psi'_n(\sqrt{\lambda}\, r))^2 
+ (\alpha_n^2 - \lambda r^2)\psi_n^2(\sqrt{\lambda}\,r) \right) , 
\end{eqnarray*}
where we used (\ref{eq:Cip_Besselnorm}) and the condition
$\psi_n(\sqrt{\lambda}\, R_2) = 0$.  We get then
\begin{equation}
\begin{split}
- \frac{\psi'_n(\sqrt{\lambda}\, r)}{\psi_n(\sqrt{\lambda}\, r)} 
& \geq  \frac{b(R_2)\, r}{2\lambda^{3/2}} 
\left(- \left(\frac{\psi'_n(\sqrt{\lambda}\, r)}{\psi_n(\sqrt{\lambda}\, r)}\right)^2 + b(r) \right), \\
\end{split}
\end{equation}
where we dropped the positive term $(R_2 \psi'_n(\sqrt{\lambda}\,
R_2))^2$.  Denoting $w = - \frac{\psi'_n(\sqrt{\lambda}\,
r)}{\psi_n(\sqrt{\lambda}\, r)}$ and $a = r b(R_2)/(2\lambda^{3/2})$,
the above inequality can be written as
\begin{equation}
a w^2 + w - ab(r) \geq 0 .
\end{equation}
Since $a > 0$ and $w > 0$, this inequality implies
\begin{equation}
- \frac{\psi'_n(\sqrt{\lambda}\, r)}{\psi_n(\sqrt{\lambda}\, r)} \geq \frac{2ab(r)}{1 + \sqrt{1 + 4a^2b(r)}}  \qquad \forall~ 0 < r < R_2,
\end{equation}
which is equivalent to (\ref{eq:psin_psi_general}).  \qed

\begin{corollary}  \label{lem:psin_positive}
If $\alpha_n^2/R_2^2 > \lambda$, then $\psi_n(\sqrt{\lambda}\, r)$ is a
positive monotonously decreasing function on the interval $(0,R_2)$:
\begin{equation}
\psi_n(\sqrt{\lambda}\, r) \geq 0 , \qquad \psi'_n(\sqrt{\lambda}\, r) \leq 0,  \qquad \forall~ 0 < r \leq R_2 .
\end{equation}
\end{corollary} \\
{\bf Proof.}  The inequality (\ref{eq:psin_ineq}) implies that
$\psi_n(\sqrt{\lambda}\, r)$ is either positive monotonously
decreasing or negative monotonously increasing on the interval
$(0,R_2)$.  We compute then
\begin{equation}
\psi'_n(\sqrt{\lambda}\, R_2) = J'_{\alpha_n}(\sqrt{\lambda}\,R_2) Y_{\alpha_n}(\sqrt{\lambda}\, R_2) 
- Y'_{\alpha_n}(\sqrt{\lambda}\, R_2) J_{\alpha_n}(\sqrt{\lambda}\, R_2)  = - \frac{2}{\pi \sqrt{\lambda}\, R_2} < 0 ,
\end{equation}
where we used the Wronskian for Bessel functions.  As a consequence,
$\psi'_n$ is negative in a vicinity of $R_2$ and thus on the whole
interval.  \qed

\begin{corollary}
If $\lambda$ is fixed by (\ref{eq:Cip_lambda}), then there exists
$\alpha_1$ large enough such that for any $n \geq 2$,
\begin{equation}
\label{eq:psin_positive}
\psi_n(\sqrt{\lambda}\, r) \geq 0 , \qquad \psi'_n(\sqrt{\lambda}\, r) \leq 0,  \qquad \forall~ 0 < r \leq R_2 ,
\end{equation}
and
\begin{equation}
\label{eq:psi2_J0}
- \frac{\psi'_n(\sqrt{\lambda}\, R_1)}{\psi_n(\sqrt{\lambda}\, R_1)} \geq - \frac{J'_0(\sqrt{\lambda}\, R_1)}{J_0(\sqrt{\lambda}\, R_1)}  .
\end{equation}
\end{corollary}  \\
{\bf Proof.}  When $\lambda$ is given by (\ref{eq:Cip_lambda}), one
has
\begin{equation}
b(R_2) = \frac{n^2 \alpha_1^2}{R_2^2} - \lambda = c_2 \lambda ,  \qquad c_2 = \frac{n^2}{(1+\ve_1)^2} - 1 > 0
\end{equation}
and
\begin{equation}
b(R_1) = \frac{n^2 \alpha_1^2}{R_1^2} - \lambda = (R_2^2 c_1 - 1) \lambda ,  \qquad c_1 = \frac{n^2}{R_1^2(1+\ve_1)^2} > 0
\end{equation}
for any $n \geq 2$ and $\alpha_1$ large enough (that makes $\ve_1$ small
enough).  As a consequence, Corollary \ref{lem:psin_positive} implies
(\ref{eq:psin_positive}), while Lemma \ref{lem:psin2} implies
\begin{equation}
\label{eq:Cip_auxil3}
- \frac{\psi'_n(\sqrt{\lambda}\, R_1)}{\psi_n(\sqrt{\lambda}\, R_1)} \geq 
\frac{R_1 \, b(R_1) \, b(R_2)}{\lambda^{3/2} + \sqrt{\lambda^3 + R_1^2 \, b(R_1) \, b^2(R_2)}} = C_n R_2 ,
\end{equation}
with
\begin{equation}
C_n = \frac{R_1 (c_1 - 1/R_2^2) c_2 \sqrt{\lambda}}{1/R_2 + \sqrt{1/R_2^2 + R_1^2 (c_1 - 1/R_2^2) c_2^2}} \longrightarrow 
\sqrt{c_1 \lambda}  \qquad \mbox{ as } R_2\to\infty .
\end{equation}
In other words, when $\lambda$ is fixed by (\ref{eq:Cip_lambda}), the
left-hand side of (\ref{eq:Cip_auxil3}) can be made arbitrarily large.
On the other hand, for a fixed $\lambda$, the term
$\frac{J'_0(\sqrt{\lambda}\, R_1)}{J_0(\sqrt{\lambda}\, R_1)}$ in the
inequality (\ref{eq:psi2_J0}) is independent of $\alpha_1$ or $R_2$.
We conclude that the inequality (\ref{eq:psi2_J0}) is fulfilled for
$\alpha_1$ (and $R_2$) large enough. \qed

\begin{lemma}
If $\lambda$ is fixed by (\ref{eq:Cip_lambda}), then there exists
$\alpha_1$ large enough such that for any $n_2 > n_1 \geq 2$, one has
\begin{equation}
\label{eq:psi_ineq}
\frac{\psi'_{n_1}(\sqrt{\lambda}\, r)}{\psi_{n_1}(\sqrt{\lambda}\, r)} \geq \frac{\psi'_{n_2}(\sqrt{\lambda}\, r)}{\psi_{n_2}(\sqrt{\lambda}\, r)} 
 \qquad \forall~ 0 < r < R_2.
\end{equation}
\end{lemma}  \\
{\bf Proof.}  Integration of (\ref{eq:J_ineq_aux}) from $r$ to $R_2$
yields
\begin{eqnarray} \nonumber
&& - r \left(\psi_{n_1}(\sqrt{\lambda}\, r) \psi'_{n_2}(\sqrt{\lambda}\, r) - \psi_{n_2}(\sqrt{\lambda}\, r) \psi'_{n_1}(\sqrt{\lambda}\, r)\right) \\
&& = \int\limits_{r}^{R_2} \frac{(\pi n_2/\phi_1)^2 - (\pi n_1/\phi_1)^2}{r'} \psi_{n_1}(\sqrt{\lambda}\, r') \psi_{n_2}(\sqrt{\lambda}\, r') dr' , 
\end{eqnarray}
where the upper limit at $r = R_2$ vanished due to the boundary
condition $\psi_n(\sqrt{\lambda}\, R_2) = 0$.  Since
$\psi_n(\sqrt{\lambda}\, r) \geq 0$ over $r \in (0,R_2)$ according to
(\ref{eq:psin_positive}), the integral is positive that implies
(\ref{eq:psi_ineq}).  \qed

Note that the sign of inequality is opposite here as compared to the
inequality (\ref{eq:J_ineq1}).

\end{document}